\documentclass[fleqn,usenatbib]{mnras}

\usepackage[T1]{fontenc}

\DeclareRobustCommand{\VAN}[3]{#2}
\let\VANthebibliography\thebibliography
\def\thebibliography{\DeclareRobustCommand{\VAN}[3]{##3}\VANthebibliography}

\usepackage{graphicx}	
\usepackage{amsmath}	
\usepackage{lscape}
\usepackage[usenames]{color}

\title[20 years of PM1-188]{Twenty years of observations of PM 1-188: Its chemical abundances and extraordinary kinematics}

\author[M. Pe\~na et al.]{
Miriam Pe\~na,$^{1}$\thanks{E-mail: miriam@astro.unam.mx (MP)}
Liliana Hern\'andez-Mart\'inez,$^{2,3}$
Francisco Ruiz-Escobedo$^{1}$
\\
$^{1}$Instituto de Astronom\'ia, Universidad Nacional Aut\'onoma de M\'exico, Ap. 70-264, Ciudad Universitaria, 04510, Ciudad de M\'exico\\
$^{2}$Instituto de Ciencias Nucleares, Universidad Nacional Autonoma de M\'exico,
 Ap. 70-543, 04510, Ciudad de M\'exico\\
$^{3}$Facultad de Ciencias, Universidad Nacional Autonoma de M\'exico, 04510, Ciudad de M\'exico
 }

\date{Accepted 2021 February 17. Received 2021 February 15; in original form 2020 September 14 }

\pubyear{2015}

\begin{document}
\label{firstpage}
\pagerange{\pageref{firstpage}--\pageref{lastpage}}
\maketitle

\begin{abstract}
The analysis of 20 years of spectrophotometric data of the double shell planetary nebula PM\,1-188 is presented, aiming to determine the time evolution 
of the emission lines and the physical conditions  of the nebula, as a consequence of the systematic fading of its [WC\,10] central star whose 
brightness has declined  by about 10 mag in the past 40 years.  Our main results include that the [\ion{O}{iii}], [\ion{O}{ii}], [\ion{N}{ii}] line intensities  are increasing with time in the inner nebula as a consequence of an  increase in electron temperature  from 11,000 K in 2005 to more than 14,000 K in 2018, due to shocks. The intensity of the  same lines are decreasing in the outer nebula, due to a decrease in temperature, from 13,000 K to 7,000 K, in the same period.
The chemical composition of the inner and outer shells was derived and they are similar. Both nebulae present subsolar O, S and Ar abundances, while they are He, N and Ne rich. For the outer nebula the values are 12+log He/H= 11.13$\pm$0.05, 12+log O/H = 8.04$\pm$0.04, 12+log N/H= 7.87$\pm$0.06, 12+log S/H = 7.18$\pm$0.10 and 12+log Ar = 5.33$\pm$0.16. The O, S and Ar abundances are several times lower than the average values found in disc non-Type I PNe, and are reminiscent of some halo PNe. From high resolution spectra, an outflow in the N-S direction was found in the inner zone. Position-velocity diagrams show that the outflow expands at velocities in the $-$150 to 100 km s$^{-1}$ range, and both shells have expansion velocities of about 40 km s$^{-1}$.
\end{abstract}

\begin{keywords}
planetary nebulae: individual: PM\,1-188 -- ISM: abundances -- ISM: kinematics and dynamics
\end{keywords}



\section{Introduction}

Planetary nebulae (PNe) are  formed from stars  of low-intermediate masses (1--8 M$_\odot$) in an advanced stage of evolution (post AGB). They have completely burnt hydrogen and helium in their cores and possess dense carbon-oxygen cores of about 0.6 M$_\odot$. At these late  stages, the star loses part of its atmosphere through intense winds. It evolves towards larger effective temperatures and starts ionising the ejected envelope. Thus the PN was part of the stellar atmosphere and its chemical composition depends on the processes of dredge-ups that carry elements of the interior to the surface. 

 In general, central stars of PNe possess a thin H-rich envelope, but about 15\% of them are known to be H-deficient, showing atmospheric instabilities and developing strong stellar winds similar to the massive Wolf-Rayet stars. They have been named [WR], to differentiate them from the massive WR. Almost all of these stars are of the sequence of Carbon, showing strong helium, carbon and oxygen emission lines,  therefore they are designated as  [WC] stars. It is found that most of them are distributed in the early spectral types [WC\,2--4] which show lines of \ion{C}{iv}, \ion{He}{ii} and \ion{O}{v} and have effective temperatures larger than  100,000 K, and in the late spectral types [WC\,8 -- 11], showing lines of lower ions and  much lower effective temperatures, from 25,000 K to 80,000 K (\citealt{koesterke:01}; \citealt{acker:03}). There are very few stars in the intermediate classes. In the following the associated nebula will be called a [WC]PN. 

PN G012.2+04.9, with common names PM\,1-188, HuBi\,1 and IRAS\,17514-1555, is a well known quite interesting planetary nebula first found by the IRAS satellite, which during several decades has shown a central star of the rare [WC\,10] class. This was discovered  by \citet{hubi:90} who also reported a strong infrared excess in this object. Since this discovery, PM1-188 has been
actively studied to reveal its nature and evolution. 

\citet{pollacco:94} found that PM\,1-188 consists of a faint extended nebula surrounding a dense bright zone, with apparently bipolar structure.  \citet{pena:05b} reported the extraordinary  ionisation structure of the nebula, showing that the [\ion{N}{ii}] and [\ion{O}{iii}] emission lines  appear concentrated in the inner zone, while the H lines  are more intense in the outer zone, presenting a cavity in the centre (see her Fig. 1). Also in this paper, the author indicated the systematic fading of the central star.

\citet{pena:05b} reported that the [WC] central star of PM\,1-188 shows a systematic fading. The WR phenomenon is known to be variable. Indeed, a few [WC] PNe show variations that could be intrinsic to the star or due to external causes (\citealt{werner:92}; \citealt{pena:97}).

 \citet{leuenhagen:98}  derived physical parameters of the central star of PM\,1-188 from  non-LTE theoretical stellar atmosphere models  finding  a mass loss rate log \.M=$-$5.70, a terminal wind velocity of about 360 km s$^{-1}$, a surface temperature of 35,000 K,  a luminosity 
 log L/L$_\odot$=3.70  and  a chemical composition, in mass fraction, of $\beta$(He) = 42\%, $\beta$(C) = 50\%, $\beta$(O) = 7\%, and a very small amount of H.  The remarkable He- and C-rich and H-poor photospheric chemical composition is typical of [WC] stars \citep{koesterke:01} and it has been found in some born-again PNe such as Abell 30, although in this case, $\beta$(He) amounts 63\%,  $\beta$(C), 20\% and $\beta$(O), 15\%, by mass \citep{todt:15}, chemistry more typical of a [WC-early] star. PM\,1-188 might have experienced an unusual evolution as Abell 30.

 \citet{pena:01} and \citet{gorny:01} analysed some nebular and dust properties in a number of [WC]PNe, and they suggest a possible evolutionary sequence from [WC-late] to [WC-early] objects, except for  PM\,1-188 and K\,2-16, both with a very late [WC] star. \citet{pena:01} suggested that these exceptions could have experienced a late-helium-flash and have returned to the AGB for a born-again evolution, or they could be stars evolving very slowly from the AGB due to their low mass. 

Recently,  \citet{guerrero:18} deeply analysed  the nebular structure and the evolution of the central star, claiming  that the inner nebula is ionised by shocks with velocity of about 70 km s$^{-1}$; the outer nebula is recombining and, according to models of stellar evolution, the central star is the descendent of a low-mass star that experienced a  born-again event whose ejecta shock-excite the inner shell.
  
   In summary of these prior works, PM1-188 is definitely a [WC 10] PN exhibiting time variations
which might be related to unusual stellar evolution. While the central star seems to be more deeply
understood, the nature of the nebula remains unclear. To further advance understanding of PM\,1-188 and gain insights on evolution of born-again phenomenon, time temporal evolution of the nebular
parameters and internal gas motion are necessary. For that purpose, we gathered 20 years of low and
high-resolution spectroscopic data and also newly secured high-dispersion spectra to perform
spatially-resolved studies of the nebula.

This paper is organized as follows: In \S 2 the observations are described. In \S 3, line intensities and their time evolution in the inner and outer nebulae  are analysed. In \S 4, we present the physical conditions (electron temperatures and densities) and their time evolution, and the derived ionic and chemical abundances for both, inner and outer nebulae. In \S 5 we discuss the evidence for shocks in the inner shell. The presence of an outflow, derived from high-resolution spectra,  is shown in \S 6, while in \S 7  Position-Velocity diagrams constructed from data obtained in two position angles, are presented. The discussion and conclusions are displayed in \S 8.  In an appendix we present the Ionisation Corrections Factors (ICFs) used to derive the chemical abundances and the list of atomic parameters employed to determine the physical conditions and chemistry of the nebula.

\section{Observations}

As said above, more than 20 years of spectroscopic observations of PM\,1-188 have been gathered. Data prior to 2005 were already published in \citet{pena:01} and \citet{pena:05a}, and are reanalysed here. New observations reported in this work include spectrophotometric data obtained in 2005 with the Las Campanas Observatory (LCO) Clay telescope and the LDSS3-Two spectrograph; observations with the OAN-SPM 2-m telescope and the Boller \& Chivens (BCh) spectrograph in 2005 and 2017, and observations with the Manchester Echelle Spectrograph (MES) in 2017. Also observations with the Gran Telescopio Canarias (GTC) and OSIRIS  spectrograph in long slit mode in  2018 (program GTC112-18A, P.I. M. Guerrero), retrieved from the public archives, are analysed here. 

  The log and characteristics of the observations are presented in Table~\ref{tab:observations_table}  where we list the telescope and instrument used, the covered  wavelength range, the spectral resolution (in \AA/pix) and resolving power R  ($\lambda$ / $\delta\lambda$), the spatial scale, the total exposure time, the slit size (length and width),  the position angle of the slit, P.A., the seeing during the observations and some comments. 
\begin{table*}
\setlength\tabcolsep{4pt}
\centering
	\caption{Log of observations and their characteristics}
	\label{tab:observations_table}
\begin{tabular}{lllccrcrll}
\hline
Date & telescope & instrument & $\lambda$ range  & \multicolumn{2}{c}{\underline{spectral resolution}}  & plate scale &total exp & slit size  & 
P.A., seeing, comments\\
         &                  &                     & \AA                      & \AA/pix & R (5000\AA)&arcsec/pix &time (s) &\\
         \hline
1997/08/04$^a$ & OAN-2m& echelle REOSC &  3600--6800&0.3&18000 & 1.20&1800 &  13.3''$\times$2''& 90$\degr$, 1.5'',  \citet{pena:01}\\
2000/09/25-26$^a$ & OAN-2m & BCh-600 l/mm$^b$ &4400--6750  & 1.54& 685 & 1.50&2400 &5$'$$\times$4''&90$\degr$, 2'', \citet{pena:05a}\\ 
2002/06/05-07$^a$ &   OAN-2m & BCh-300 l/mm & 3450--7400 &  2.23&567 &  1.50&4200  & 5'$\times$4''&90$\degr$, 1.5'', \citet{pena:05a}\\ 
2004/04/23-26$^a$ & OAN-2m & echelle-REOSC  &3800--6800 &0.3&18000& 1.20&2700 & 13.3''$\times$2'' & 90$\degr$, 2'', \citet{pena:05a}\\
2005/05/15-17 & OAN-2m & BCh-600 l/mm&3800--5950&  1.54& 685& 1.20&7200 & 5'$\times$4'' & 90$\degr$, 1.5'', \\ 
2005/05/18-19&OAN-2m&BCh-600 l/mm& 5200--7300&  1.54& 685&1.20 &3600 &  5'$\times$4''& 90$\degr$, 1.5''\\
2005/08/13 & LCO Clay& LDSS3-Two &3830--5150  & 0.8& 1900&0.38& 900 & 4'$\times$2'' & 173$\degr$ (parallactic), 1''-1.2'' \\
 ~~~~~''&~~~~~'' & ~~~~~'' & 5160--6595 & 0.8& 1900 &0.38& 900 &4'$\times$2''& 173$\degr$ (parallactic), 1''-1.2'' \\
 2017/06/24,25 &OAN-2m & BCh-300 l/mm &3700--7360 &  2.23 & 685& 1.08 &7200 & 5'$\times$2''& 90$\degr$, 1.5'' \\  
 2017/06/29& OAN-2m & MES order-87 & 6545--6595& 0.06&25000&  0.35&7200 & 5.2$'$$\times$2'' &90$\degr$, 1.5'' \\
 2017/06/30 & OAN-2m &MES order-87 & 6545--6595 & 0.06&25000&0.35&7200&  5.2$'$$\times$2'' &0$\degr$, 2.5''  \\
 2018/05/14 & GTC$^c$ & OSIRIS LS & 3630--7500  &  2.12&1018 & 0.254&6300 &  6.5$'$$\times$0.8'' & 90$\degr$, 1'' \\ 
   \hline 
   \multicolumn{8}{l}{a: published in \citet{pena:01} or in \citet{pena:05a}}\\
   \multicolumn{8}{l}{b: For the BCh spectrograph, the used grid is indicated}\\
   \multicolumn{8}{l}{c: GTC112-18A program, P.I. M. Guerrero, data retrieved from the public archives}
   \end{tabular}
   \end{table*}

\subsection{Description of spectroscopic data}

Most of the data along these years was obtained at Observatorio Astron\'omico Nacional, San Pedro M\'artir, B.C., M\'exico (OAN-SPM), with the 2.1-m telescope and the Boller \& Chivens (BCh) spectrograph, at intermediate resolution (4 -- 7 \AA/pix).  Most of the time, the observed wavelength range  was from about 3700 to 7400 \AA~ which allows to obtain a large number of lines emitted in the visual zone.  The spectral resolution of these observations is adequate to resolve the important [\ion{N}{ii}] and [\ion{S}{ii}] lines employed in temperature and density diagnostic.

In two epochs (1997 and 2004) the high resolution  echelle REOSC spectrograph, that achieves a spectral resolution of about 0.3 \AA/pix (R=18000 at 5000 \AA),  was used. In 2017, the Manchester Echelle Spectrograph (MES) was employed in order  to derive the interesting kinematics of this nebula; these observations are described in the next subsection.

In 2005 spectra with the LDSS3-Two long-slit spectrograph, attached to the LCO 6.5-m  telescope Clay were obtained.  This spectrograph works with two arms operating simultaneously. The blue one covered from  3830 to 5150 \AA~ and the red one, from 5160  to 6595 \AA. The grism VPH Blue was employed.  Standard stars  and a He-Ne-Ar lamp were used for flux and wavelength calibrations. The usual procedures with IRAF \footnote{IRAF is distributed by the National Optical Astronomy Observatories, which is operated the Association of Universities for Research in Astronomy, Inc., under contract to the National Science Foundation.} were performed  for data reduction. The nebular line fluxes were measured in the calibrated spectra, integrating all the emission over a local continuum estimated by eye.

\begin{figure}
	\includegraphics[width=\columnwidth]{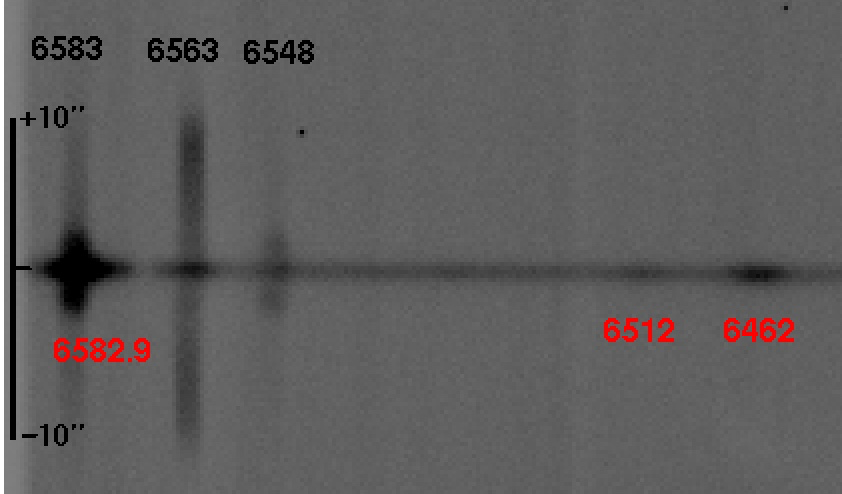}
    \caption{The  nebular lines H$\alpha$ $\lambda$6563 and [\ion{N}{ii}] $\lambda\lambda$6548,6583  are shown from the 2D LCO-Clay-LDSS3 spectrum, obtained in 2005, with the slit at P.A. 173$^\circ$. The spatial scale is indicated in the left side, with the origin in the stellar continuum. The extended nebula is clearly visible in these lines. The compact inner nebula presents intense [\ion{N}{ii}] lines, diminishing towards the outer nebula, while  H$\alpha$ is more intense in the outer nebula. The wavelengths of nebular lines are marked in black. The stellar continuum and several [WR] lines of \ion{C}{ii} are detected and marked in red. In particular, the \ion{C}{ii} $\lambda$6582.9 line, overlapped with the [\ion{N}{ii}] $\lambda$6583 line, is very intense.}
    \label{fig:star_LCO}
\end{figure}

The central star is not visible in the spectra of 2017 or 2018 due to its faintness. But it was well detected in 2005, when data with LCO telescope Clay were obtained. Fig. \ref{fig:star_LCO}  shows a zone around H$\alpha$, where the stellar continuum and the extended nebula are visible. Nebular [\ion{N}{ii}]  lines  are very intense in the inner zone, [\ion{N}{ii}]$\lambda$6583 in particular appears contaminated with the stellar \ion{C}{ii} $\lambda$6582.9 line.  [\ion{N}{ii}] lines are faint in the outer zone, but perfectly detected.  H$\alpha$  appears intense in the centre, overlapped to the stellar continuum, then it shows a cavity immediately outside and increases its emission further away in the zone of the outer nebula. The nebular lines are marked in the figure. The stellar emission shows several WR lines of \ion{C}{ii}, also marked in the figure. 

The most recent observations presented here were obtained in 2018 with the GTC OSIRIS spectrograph in long slit mode. Data were retrieved from the archives and data reduction  was performed with the standard procedures using IRAF. Flux and wavelength calibrations were carried out with the standard star GD140 and the Hg-Ar-Ne lamp provided in the same observations. 
These spectra covered a wide wavelength range from 3700 to 7300 \AA. The large exposure time (105 min in total, distributed in 4 individual observations of 1575 s each) helped to detect a large number of nebular lines, which allowed us to determine confident physical conditions and ionic abundances.   Nebular line fluxes were measured in the calibrated spectra, integrating all the emission over a local continuum estimated by eye.

\smallskip 

All the spectra of a given epoch were combined in a single spectrum to increase the signal-to-noise. The total observing time for each epoch is listed in Table \ref{tab:observations_table}. 
For the analysis of lines, we made an effort  to obtain separately the line intensities of the inner and outer nebulae in all the spectra, although in the epochs previous to 2005,  the [WC] central star was very bright and some of its emission lines are  contaminating some nebular lines such as the \ion{He}{i} $\lambda\lambda$4471,5876,6678 lines and the [\ion{N}{ii}] $\lambda$6583 line in the low spectral-resolution data. 
Besides, it was not possible  to extract the outer nebula separately, then the line fluxes presented in Table \ref{tab:outer-nebula} for 2000 and 2002, include all the nebular emission.

On the other hand, spectra obtained with LCO-Clay-LDSS3 in 2005, with  OAN-SPM-BCh in 2017 and  with  GTC-OSIRIS-LS in 2018 could be well separated in inner and outer emission, and these data are not contaminated by the stellar emission. 

In general, the  spatial profiles of lines  in all our spectra corroborate the behaviour presented by \citet{guerrero:18} regarding the emission structures of the inner and outer nebulae. Lines of the ions N$^+$ ($\lambda\lambda$5755, 6548, 6583), O$^+$ ($\lambda\lambda$3727, 7325), O$^{++}$ ($\lambda\lambda$4959, 5007), He$^{++}$ ($\lambda$4686), Ne$^{++}$ ($\lambda$3869), S$^+$ ($\lambda\lambda$6716, 6717), Ar$^{++}$ ($\lambda$7135), O$^0$ ($\lambda$6300) and N$^0$ ($\lambda$5200)  are concentrated in an inner zone of about 2.5$''$, and all but [\ion{O}{ii}]$\lambda$3727  present a small central cavity. H$\alpha$, H$\beta$, \ion{He}{i} 5876 \AA, and other H and \ion{He}{i} lines are mainly concentrated in the outer nebula, and their distribution show a central cavity of about 2.5'' in the zone where low-ionisation lines are intense.

\begin{figure*}
	{\includegraphics[width=\columnwidth]{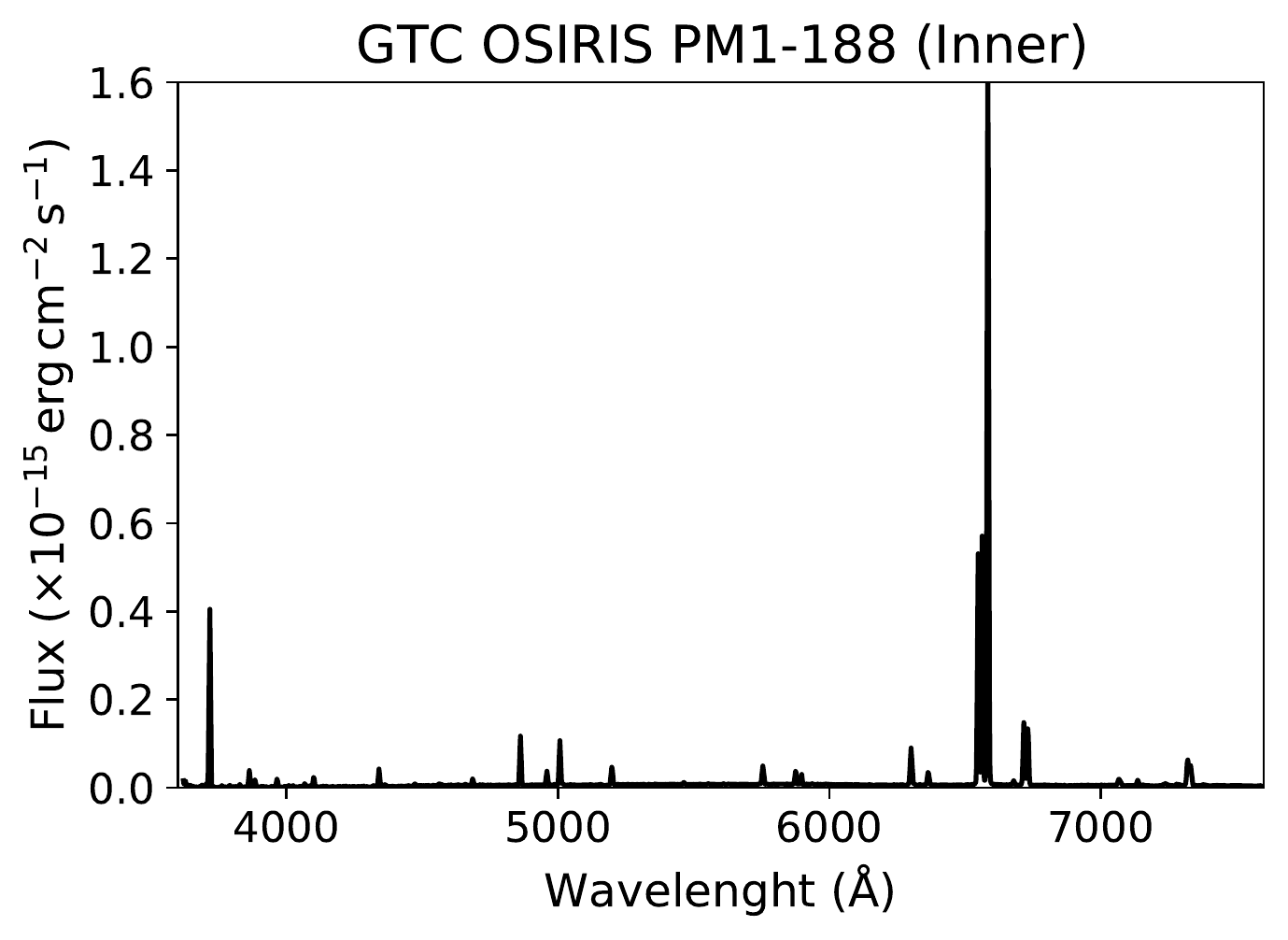}}
	{\includegraphics[width=\columnwidth]{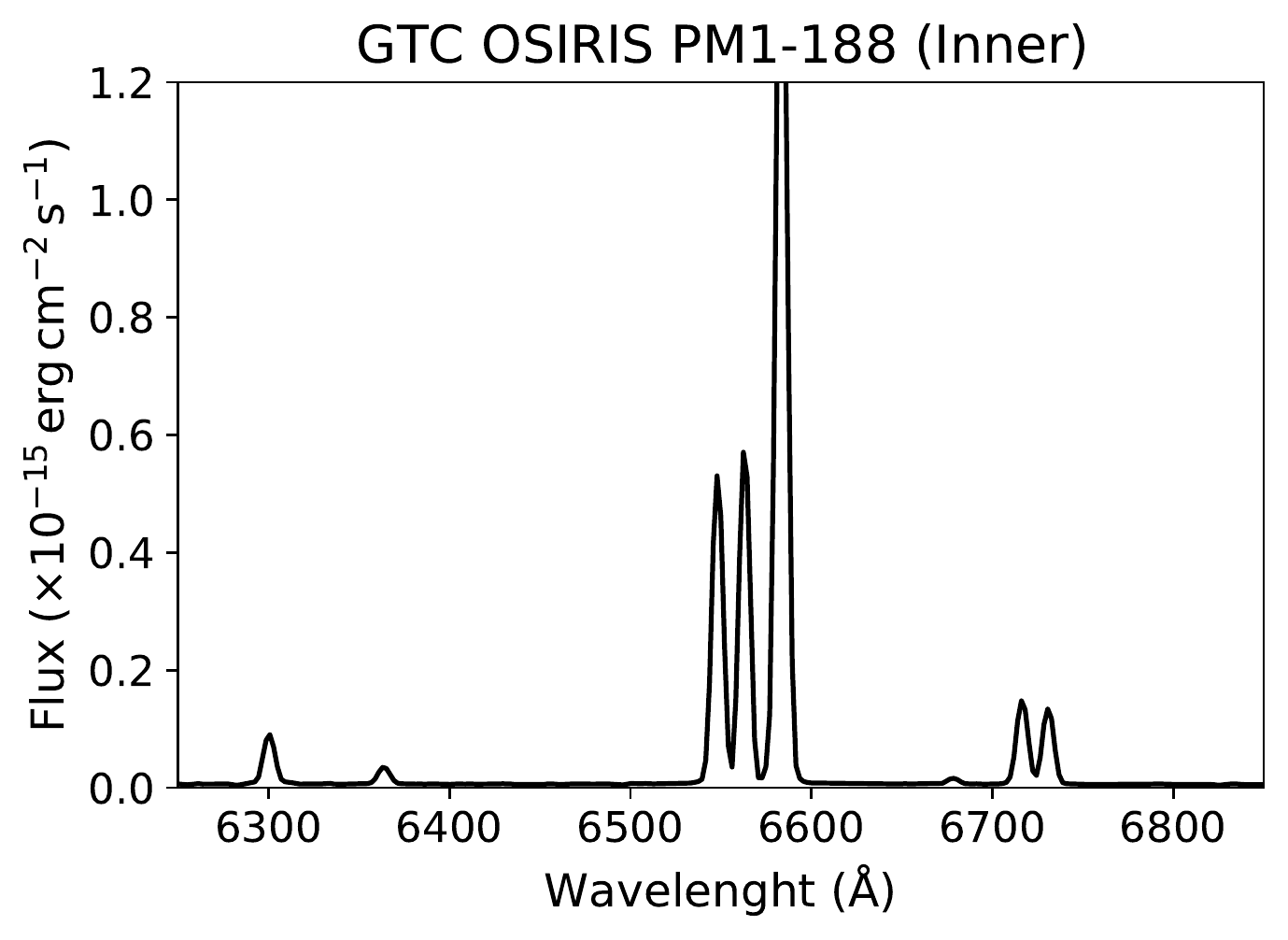}}
	{\includegraphics[width=\columnwidth]{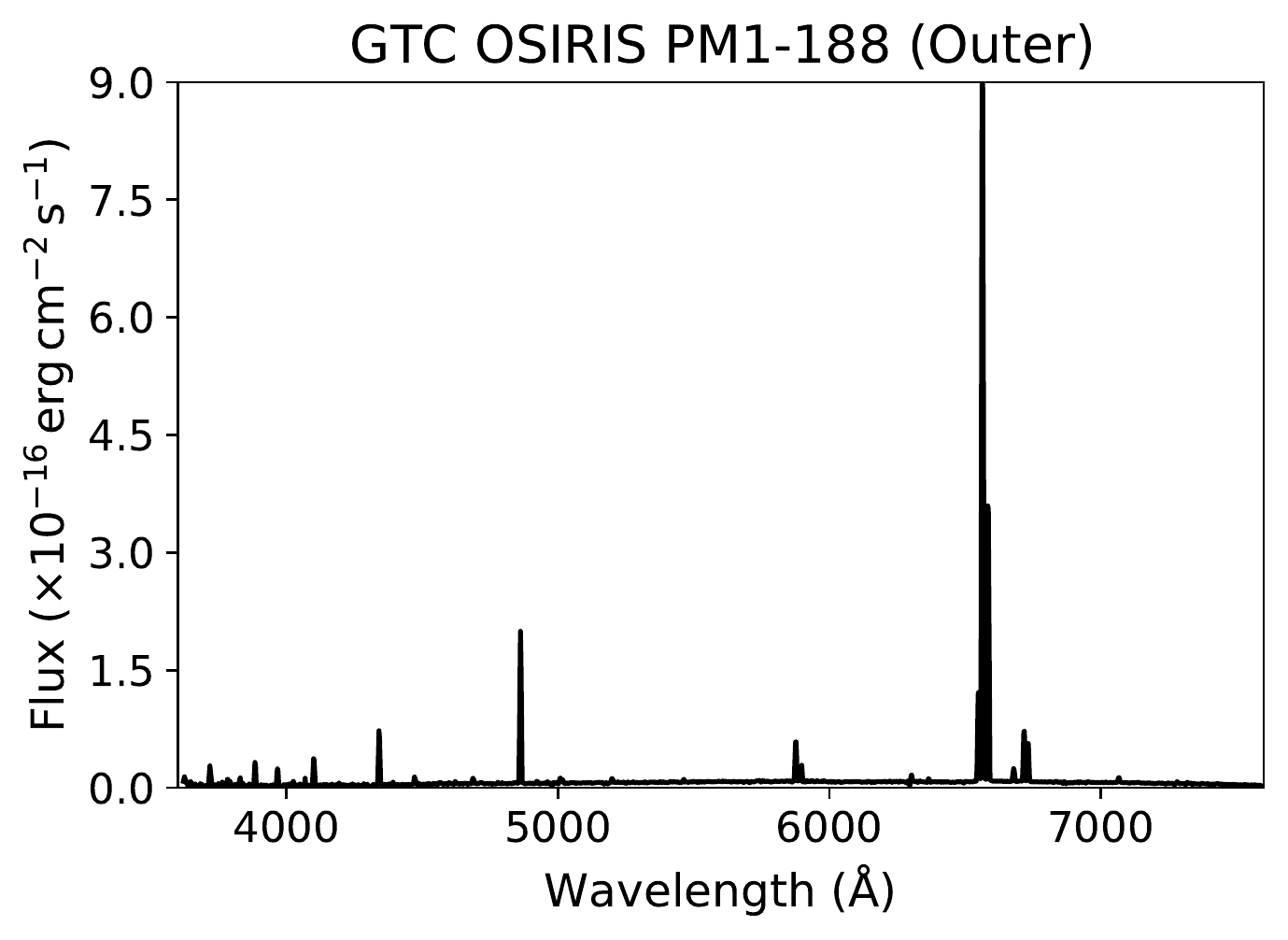}}
	{\includegraphics[width=\columnwidth]{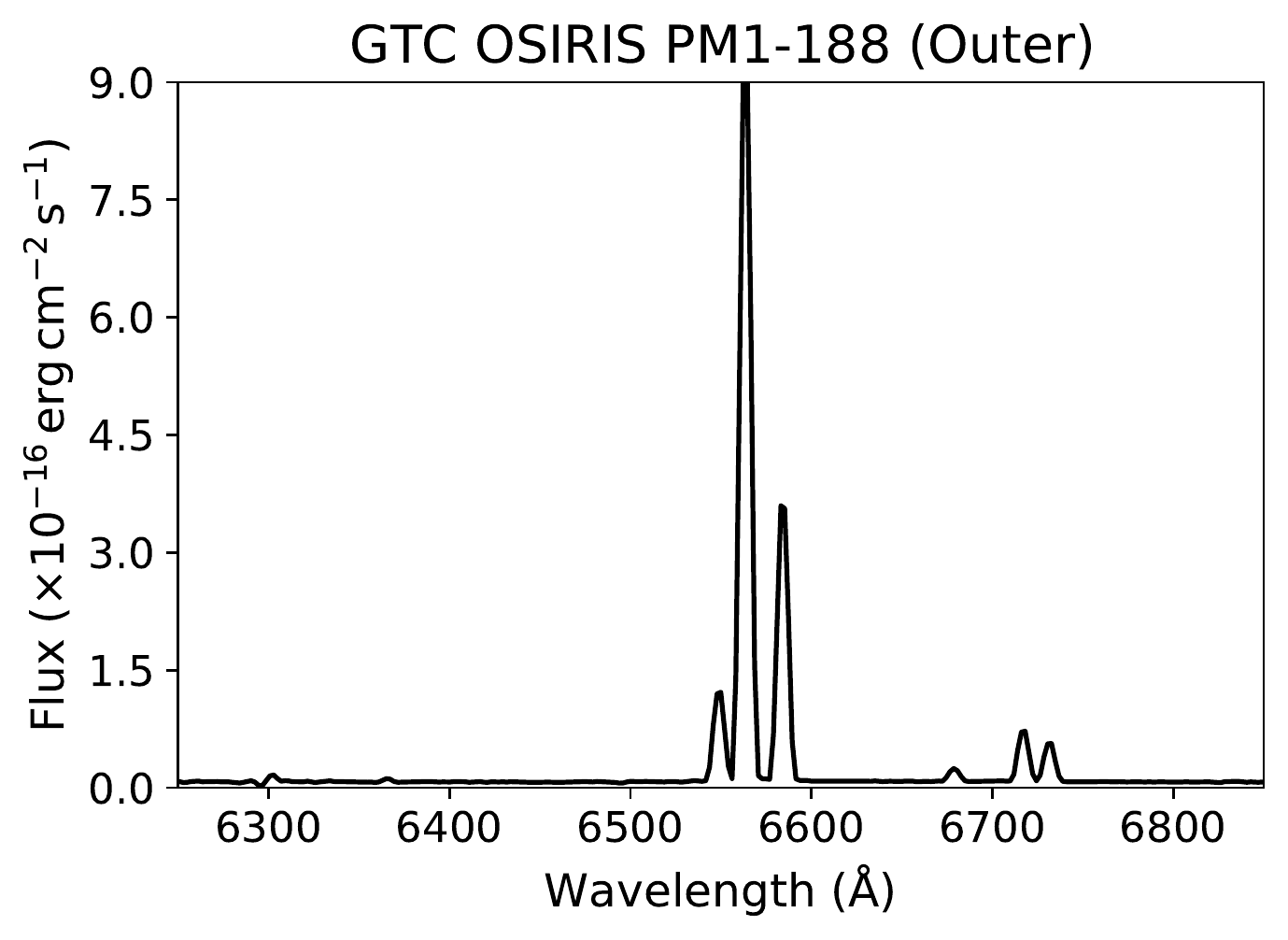}}
    \caption{The calibrated spectra of the inner (up) and outer nebulae (down), obtained from GTC OSIRIS observations, are shown. Details around H$\alpha$ in order to notice the behaviour of [\ion{N}{ii}] lines are shown on the right side.}
    \label{fig:spectra}
\end{figure*}

 Calibrated extracted spectra for both nebulae, as obtained from GTC OSIRIS data, are presented in Fig. \ref{fig:spectra}, where the different behaviour of [\ion{N}{ii}] $\lambda\lambda$6548,6583 lines, relative to H$\alpha$, in the inner and outer nebula, is noticeable.

\subsection{ MES observations}

  High resolution spectra were obtained with the OAN 2.1m telescope and the Manchester Echelle Spectrometer (MES, \citealp{meaburn:84}; \citealp{meaburn:03}) on 2017-06-29.
 Four spectra of  1800 s exposure time each were acquired, with the slit oriented E-W. These spectra were combined in one of 7200 s of total exposure time. The slit was 2$''$ wide and 6.5$'$ length. A $\delta \lambda$ = 90 \AA ~bandwidth filter was used to isolate the order 87, covering the interval from 6545 \AA ~to 6595 \AA, which then includes H$\alpha$ and [\ion{N}{ii}] 6548 \AA~ and 6583 \AA ~lines. A  binning of 2$\times$2 was implemented giving a spectral resolution of 0.06 \AA/pix, equivalent to 11  km s$^{-1}$, and a spatial resolution of 0.35 $''$/pix. The wavelength calibration was performed with a Th-Ar lamp, no flux calibration was done. 
 
  Another spectrum was obtained with MES on 2017-06-30, with total exposure time of 3600 s, and the slit oriented N-S (P.A. 0$\degr$). On this and the other occasions the slit was centred in the central star position.

\section{Line fluxes and time variation of lines in the inner and outer nebulae}

Emission of the inner nebula was extracted from the two central arcsecs. 
Fluxes  measured for this zone,  for the different  epochs, relative to H$\beta$,  not corrected by reddening, F($\lambda$),  are presented in Table \ref{tab:inner-nebula}. 
 The latter two lines in this table present, for each epoch, the total flux measured for H$\beta$ through the slit and the logarithmic reddening correction, c(H$\beta$) with its uncertainties. For each observation, the logarithmic reddening correction  was derived by using \citet{cardelli:89} reddening law by assuming a ratio of total to selective extinction $\mathrm{R_{V} = 3.1}$ and by using the  theoretical H$\alpha$/H$\beta$ ratios given by \citet{storey:95} for temperatures of 7,500 K, 10,000 K, 12,500 K and 15,000 K, adequate to the value derived for each observation. For all cases,  case B recombination theory and a density of 100 $\mathrm{cm^{-3}}$ were assumed. In the calculus of c(H$\beta$) we did not take into account the minimal contamination of \ion{He}{ii} $\lambda$4860 and $\lambda$6560 in the intensities of H$\beta$ and H$\alpha$ which in both cases amount less than 0.5\%, for the inner nebula.

The observed fluxes were dereddened by using the mentioned reddening law, and the c(H$\beta$) value determined in each case. The results, relative to I(H$\beta$), are presented in the same table, under the header I($\lambda$). The uncertainties for I($\lambda$), calculated by considering the uncertainties in the fluxes and in c(H$\beta$), are presented in \%, enclosed in parenthesis.

The emission of the outer nebula was extracted from the zones  around the central cavity presented by  the H lines. Both zones at each side of the centre were extracted and then combined. The fluxes F($\lambda$), relative to H$\beta$, not corrected by reddening, are presented in Table \ref{tab:outer-nebula}. The same as in the case of the inner nebula, fluxes were dereddened with  \citet{cardelli:89} reddening law and c(H$\beta$) derived in each case. These intensities, relative to I(H$\beta$), and their respective uncertainties, in percentage, are listed in the  same table. The latter two lines of the table indicate the observed H$\beta$ fluxes and the derived logarithmic reddening corrections c(H$\beta$) with their uncertainties. 

 The analysis of the line intensities in Tables \ref{tab:inner-nebula} and \ref{tab:outer-nebula}
indicates that prior to 2005, the \ion{He}{ii} $\lambda$4686 line, if it existed, would appear blended (or hidden) in the WR bump at $\lambda\lambda$4640-4686. As time goes by and the central star weakens, the WR bump disappears and the nebular \ion{He}{ii} line brights alone with an intensity of about 0.1 H$\beta$ in 2017 - 2018. It could be that the line intensity has been slightly increasing with time from 0.10 in 2005  to 0.13 in 2018, relative to H$\beta$, but the uncertainties (of about 15 \% for GTC OSIRIS value and larger for OAN-BCh and LCO-LSST2 values) make this not conclusive. 
In the outer nebula \ion{He}{ii} $\lambda$4686 is not detected in the spectra of LCO (2005) nor in spectra of OAN-SPM (2017), and GTC (2018).

The low ionisation lines of [\ion{O}{ii}], [\ion{N}{ii}], and [\ion{S}{ii} are intense in the inner nebula. In particular, the [\ion{N}{ii}]$\lambda$6548 shows values about half or more H$\alpha$, indicating that the nebular [\ion{N}{ii}]$\lambda$6583 (which is about 3 times more intense than $\lambda$6548) would be several times H$\alpha$. This is indicative of possible shocks as it will be discussed in \S 5. 

The [\ion{O}{iii}]$\lambda$5007 is also intense in the inner zone, although it is always fainter than H$\beta$, as expected due to  this nebula is ionised by a 35,000 - 38,000 K effective temperature star which does not produce enough photons to ionise a large fraction of O$^+$.

In the outer nebula all these low ionisation lines are detected, but much fainter, for instance the [\ion{O}{ii}]$\lambda$3727/H$\beta$ intensity ratio changes from about 7.9  in the inner zone to 0.4 in the outer zone, and I([\ion{N}{ii}]$\lambda$6548)/I(H$\beta$) has a value of about 2.7 in the inner zone and about 0.4 in the outer nebula (both values from the dereddened line observations on 2018). [\ion{O}{iii}]$\lambda$5007 is barely detected in the outer nebula. Other lines appearing in the inner nebula, that are faint or non-existent  in the outer nebula, are [\ion{Ne}{iii}]$\lambda$3869 and [\ion{Ar}{iii}]$\lambda$7135.

\begin{figure*}
    \includegraphics[scale=0.55]{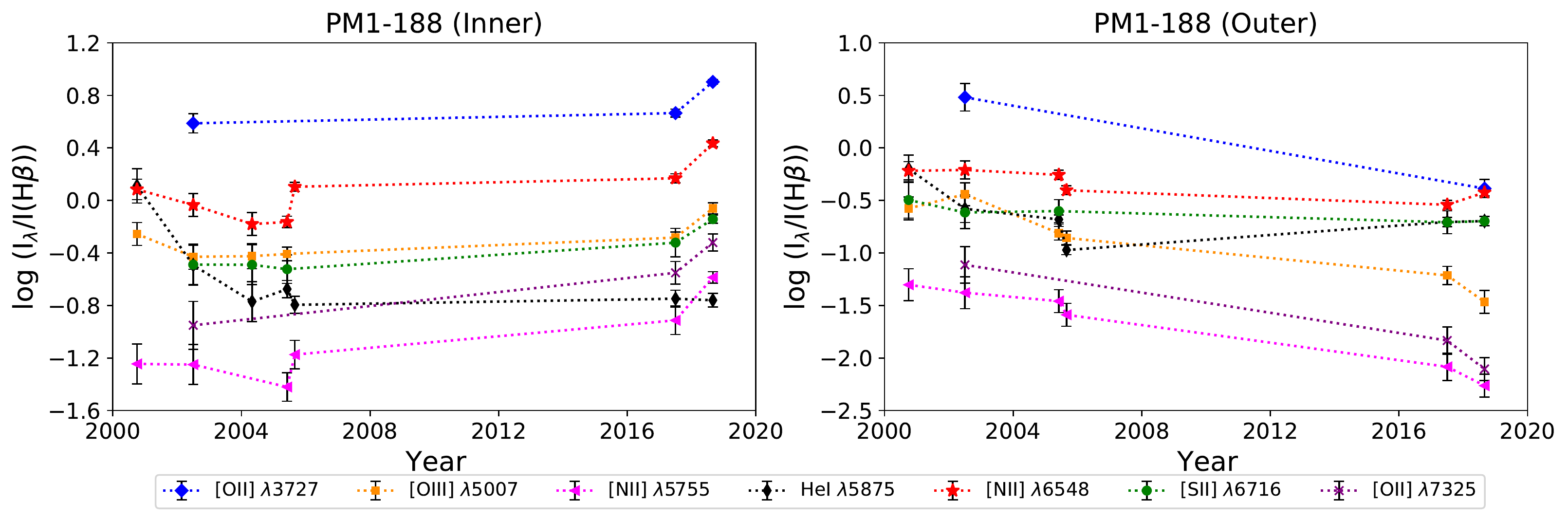}
    \caption{Time evolution of the intensity of most important lines are presented, for the inner and outer nebulae. Line intensities are presented in log-scale.}
    \label{fig:line-evolution}
\end{figure*}

 The time evolution of several lines, in the inner and outer nebulae,  are presented in Fig. \ref{fig:line-evolution}
where it is found that the [\ion{O}{iii}]$\lambda$5007, [\ion{O}{ii}]$\lambda\lambda$3727,7325, [\ion{S}{ii}]$\lambda$6716, [\ion{O}{i}] $\lambda$6300 and the [\ion{N}{ii}]$\lambda\lambda$5755,6548 lines have increased 
with time in the inner nebula, while the opposite occurs in the outer one. Such a behaviour of the collisionally excited lines seems to be a consequence of the electron temperature that increases with time in the inner nebula and decreases in the outer nebula as it is explained in the next section.

  The evolution of [\ion{N}{ii}]$\lambda\lambda$5755,6548 lines in the inner nebula, deserves a particular mention, as their intensities appear slightly declining from 2000 to 2004 and then increasing abruptly in 2005 to continue increasing up to 2018. This behaviour could be a consequence of the evolution of temperature in this nebula. The increase in temperature is affecting all the collisionally excited lines, but it is affecting in particular the [\ion{N}{ii}] lines that are very sensitive to electron temperatures and shocks. Unfortunately the [\ion{O}{ii}]$\lambda\lambda$3727,7325 lines were not observed in 2005 to verify if they show the same behaviour.

In the inner nebula the \ion{He}{i} lines are intense and they are affected by blends with stellar lines in the spectra previous to 2005.  That is the reason of the large flux of \ion{He}{i}$\lambda$5876 in 2000 - 2002. Afterwards when the stellar brightness declines, \ion{He}{i}$\lambda$5876  is not contaminated by the star and it shows an intensity near 0.2 I(H$\beta$) in both, the inner and outer nebulae. The \ion{He}{i} lines have not varied since 2005. As said before by \citep{pena:05a} \ion{He}{i} lines are intense in both nebulae, indicating a large He abundance. 

The logarithmic reddening correction, c(H$\beta$), has not changed with time, with an average value of 1.04$\pm$0.04 in the inner nebula and  0.94$\pm$0.06 in the outer nebula (considering only the  values from LCO 2005, SPM 2017, and GTC 2018, which are not affected by stellar emission). Both values are equal within uncertainties.

\section{Physical conditions and Chemical Abundances of the nebulae}

From the nebular lines, in particular from those collisionally excited lines, physical conditions such as electron densities and temperatures can be derived from some diagnostic line ratios. In the inner and outer nebulae, densities can be determined from the [\ion{S}{ii}] $\lambda\lambda$6716/6731  intensity ratios and electron temperatures from the [\ion{N}{ii}] $\lambda\lambda$5755/6548 ($\lambda$6583 line was avoided due to  possible contamination with the stellar \ion{C}{ii} $\lambda$6582.9 line), [\ion{S}{ii}] $\lambda\lambda$4086/(6716+31) and [\ion{O}{ii}] $\lambda\lambda$3727/7325 intensity  ratios.

All the densities and temperatures for different epochs, for inner and outer nebulae, were calculated with PyNeb \citep{luridiana:15}, using the atomic data presented in Table \ref{tab:atomic-parameters} (Appendix). PyNeb routine getCrossTemden was used to determine simultaneously the temperature and density from the [\ion{N}{ii}] and [\ion{S}{ii}] diagnostic lines. The results are presented in Table \ref{tab:abundance_table}. Densities derived from [\ion{S}{ii}] lines are always low, but the inner nebula is slightly denser (a few hundred particles per cm$^3$) than the outer nebula that has a density of about a hundred  particles per cm$^3$. 

Regarding the temperature, the value derived from the [\ion{N}{ii}] $\lambda\lambda5755/6748$ line ratio for the outer nebula is near 7,000 K in 2018, and more than 14,000 K in the inner nebula. Both temperatures are higher  (but very uncertain) from the [\ion{S}{ii}] $\lambda\lambda4068/(6716+31)$ and [\ion{O}{ii}] $\lambda\lambda$3727/7325 diagnostic ratios. 

\begin{figure}
	{\includegraphics[width=\columnwidth]{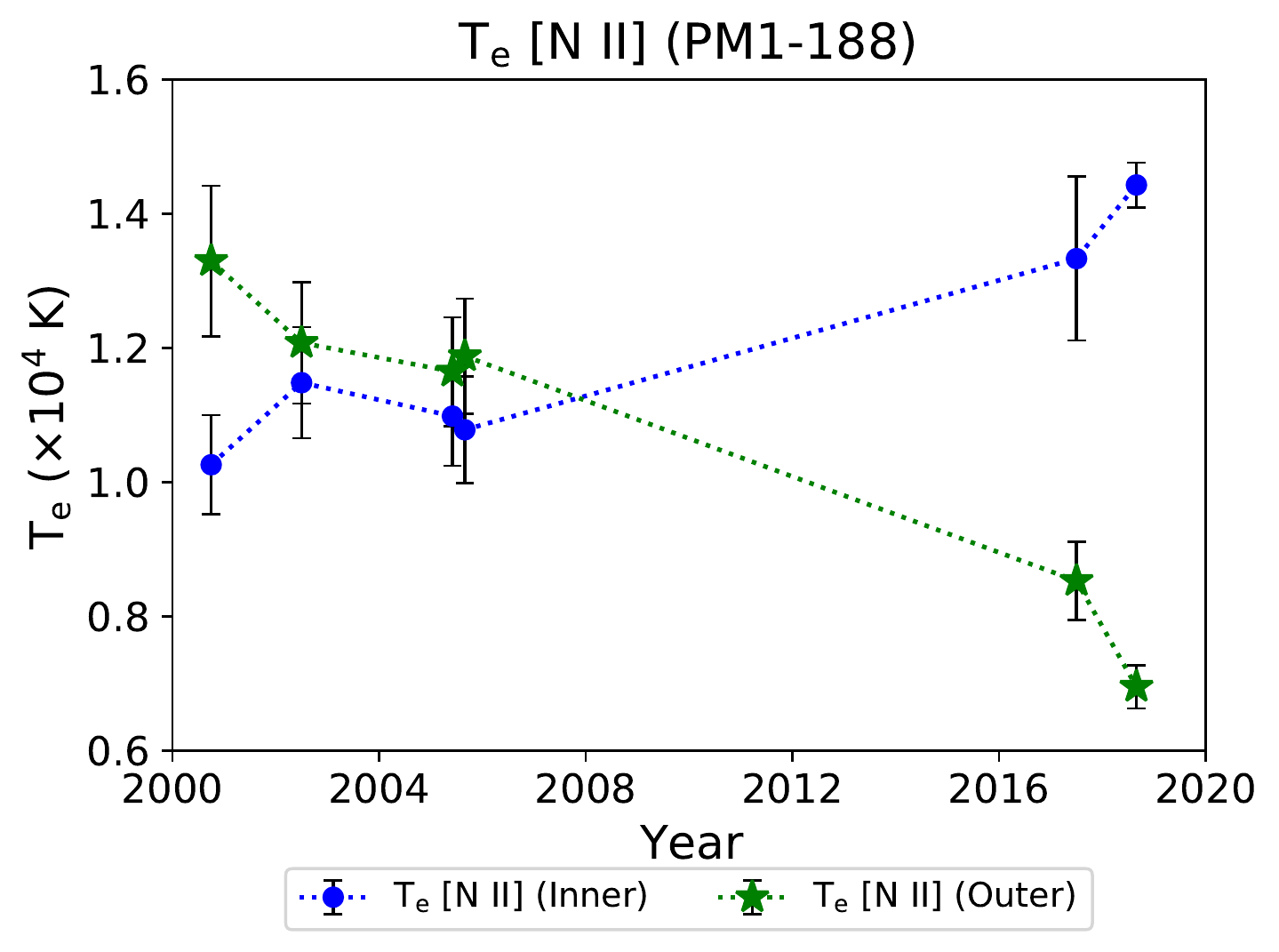}}
    \caption{Time evolution of electron temperature in the inner (blue) and outer  (green) nebulae. }
    \label{fig:te_time}
\end{figure}

 The evolution of the electron temperature in both nebulae, derived  from the [\ion{N}{ii}]  diagnostic ratios, is plotted as a function of time in Fig. \ref{fig:te_time}. Despite the errors that can be as large as 1,000 K, it is evident an opposite  systematic behavior in both nebulae. In the inner one the electron temperature has increased from about  11,000 K in the epoch 2000 - 2005 to more than 14,000 K in 2018. According to the photo-ionisation model including shock excitation, presented by \citet{guerrero:18}, this inner nebula is apparently heated by a shock with velocity  $\sim$ 70 km s$^{-1}$  traveling outwards the ionised shell. The origin of the temperature rise in this zone should be the traveling shock  and the outflow  expanding at about 150 km s$^{-1}$ in this zone (see \S 6). The rise in electron temperature is producing the increase in line intensities of heavy elements in this zone, affecting in particular the [\ion{N}{ii}] lines, as shown in Fig. \ref{fig:line-evolution}.

On the opposite, the electron temperature has decreased in the outer nebula, from about 13,000 K  in the epoch 2000-2005 to about 7,000 K in 2017-2018. This phenomenon is certainly showing the effect of the fading of the central star. Less and less ionising photons have been arriving to the external nebula since few decades ago, and as a consequence, the gas is cooling fast, and also recombining, but at a much slower rate. Supplementary Fig 4. by \citet{guerrero:18} shows a simple photo-ionisation model  evaluating the time evolution of electron temperature, the ionisation fraction of H, and the emissivities of various lines in a spherical photo-ionised nebula, after the turn off of the ionising  source. This model predicts that emissivities of [\ion{N}{ii}], [\ion{O}{ii}] and [\ion{S}{ii}] lines decrease slowly up to about 100 yr, and at this point all these emissivities decrease abruptly, when the electron temperature diminishes to less than about 7,000 K. On the other hand, the emissivity of [\ion{O}{iii}] $\lambda$5007 line decreases much faster. Our data for this zone, showing the decrease in temperature and the decay of heavy element line intensities, explained in the previous section, corroborate these predictions.

\subsection{Ionic and total abundances}

Ionic abundances were calculated by using the physical conditions derived for each epoch and the dereddened line intensities presented in Tables \ref{tab:inner-nebula} and \ref{tab:outer-nebula}. The electron density derived from [\ion{S}{ii}] $\lambda\lambda$6716/6731 line ratio and the temperature derived from the [\ion{N}{ii}] $\lambda\lambda 5755/6548$ line ratio were used in each case. When no density value was available, we assumed a value of 100 cm$^{-3}$. The results are presented in Table~\ref{tab:abundance_table}.  
\smallskip

Total abundances were derived from the ionic abundances and the ionisation correction factors (ICF) from \citet{kingsburgh:94} and \citet{delgado:14} to correct for the ions not visible in the optical spectrum. The expressions used for calculating the ICFs and to derive the total abundances are listed in the Appendix. The values obtained  in each case are presented in Table \ref{tab:abundance_table}, below the ionic abundances of each element. The derived total abundances are also listed in Table \ref{tab:abundance_table}.

In the following, for the discussion, we adopt the abundances obtained from the GTC 2018 observations for both nebulae, which present the lowest uncertainties and are more complete. PM\,1-188 outer nebula shows sub-solar O, S and Ar abundances, with  12+log O/H = 8.05$\pm$0.04, 12+log S/H = 7.18$\pm$0.10 and 12+log Ar/H = 5.33$\pm$0.16 but it is He and N rich, with 12+log He/H = 11.14$\pm$0.05 and log N/O = $-$0.18$\pm$0.19 (solar abundances as given by \citealt{asplund:09} have been adopted). This occurs as well in the inner shocked nebula  which shows 12+log O/H = 7.99$\pm$0.05,  12+log He/H = 11.13$\pm$0.05, log N/O = $-$0.06$\pm$0.07, log Ne/O = 0.26$\pm$0.08, log S/O = $-$0.99$\pm$0.11, and log Ar/O = $-$2.32$\pm$0.09. 
Thus, in addition to be He and N rich, the inner zone  appears very Ne enriched, relative to O.  Neon  cannot be measured in the outer zone, due to the absence or faintness of Ne lines. It is found that the chemical abundances of He, O, S, and Ar in the inner and outer nebulae are equal considering the uncertainties. In \S8, we discuss the chemistry in PM\,1-188, in comparison with other PNe in the galaxy.

\defcitealias{delgado:14}{DI14} 
\defcitealias{kingsburgh:94}{KB94}

\section{Shocks in PM 1-188? Diagnostic diagrams}

 BPT diagnostic diagrams \citep{baldwin:81,veilleux:87} 
have been used for many years to analyse the ionising mechanism producing an ionised region. In fact, \citet{baldwin:81}, 
suggested a two-dimensional quantitative classification scheme to separate, verbatim,  ``normal \ion{H}{ii} regions, planetary nebulae, objects photoionised 
by a power-law continuum, and objects excited by shock-wave heating’’. The most used line intensity ratios are log [\ion{O}{iii}]$\lambda$5007/H$\beta$ vs. log [\ion{N}{ii}]$\lambda$6583/H$\alpha$, log [\ion{O}{iii}]$\lambda$5007/H$\beta$ vs. log [\ion{S}{ii}]($\lambda\lambda$6717+31)/H$\alpha$, and log [\ion{O}{iii}]$\lambda$5007/H$\beta$ vs. log [\ion{O}{i}]$\lambda$6300/H$\alpha$. 

More recently \citet{kewley:01} built a theoretical classification scheme to establish an upper limit for starburst models on the optical BTP diagnostic diagrams. 
While BPT diagrams were constructed to separate  nebulae with different ionisation mechanisms and the theoretical \citet{kewley:01} models where constructed to separate starbursts from AGNs, the physics of the line formation in PNe is the same, therefore we will  study the nature of the line mechanism production in the inner and outer nebulae in PM1-188 through these diagrams. This is not the first time that  diagnostic diagrams are used to this purpose in PNe, (see e.g., \citealt{raga:08}; \citealt{akras:16}, and references therein). 

\begin{figure*}
	\includegraphics[width=18cm]{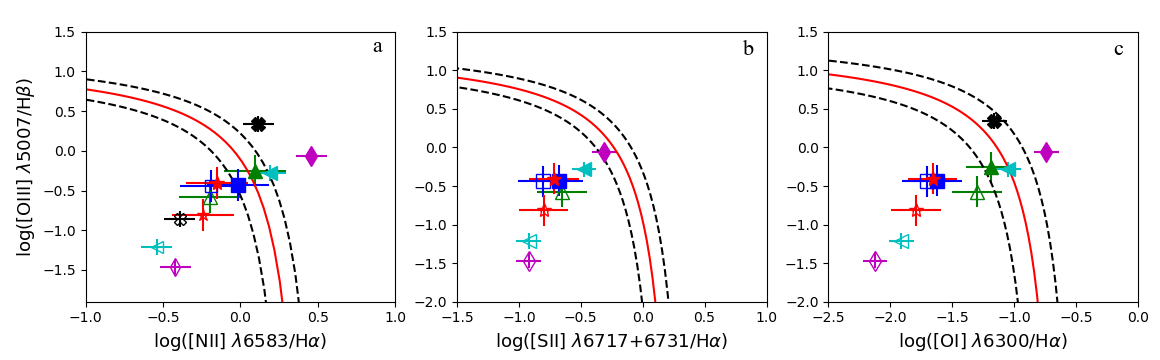}
    \caption{BPT diagrams, showing the behaviour of [\ion{O}{iii}]$\lambda$5007/H$\beta$ vs. [\ion{N}{ii}]$\lambda$6584, [\ion{S}{ii}]$\lambda$(6716+6731), and [\ion{O}{i}]$\lambda$6300, relative to H$\alpha$. The solid and dashed lines in this figure correspond to \citet{kewley:01} models and their uncertainties, respectively, and separate  \ion{H}{ii} region-like objects from AGNs. The different epochs are marked with different symbols, OAN-2000: vertical triagles, OAN-2002: squares, OAN-2005: stars, LCO-2005: crosses, OAN-2017: horizontal triangles, GTC-2018: diamonds. The filled symbols correspond to the inner nebula and the empty symbols, to the outer zone. The points to the right of the solid line are interpreted as regions heated by shocks.}
    \label{fig:bpt}
\end{figure*}

In Fig. \ref{fig:bpt}, we present diagrams of [\ion{O}{iii}]$\lambda$5007/H$\beta$ vs. [\ion{N}{ii}]$\lambda$6583/H$\alpha$,  [\ion{S}{ii}]$\lambda$(6717+6731)/H$\alpha$, and [\ion{O}{i}]$\lambda$6300/H$\alpha$ for the inner and outer nebulae. The dereddened intensity values are  taken from Tables \ref{tab:inner-nebula} and \ref{tab:outer-nebula} and are plotted with  filled  and open  symbols for the inner and outer nebulae, respectively. The different epochs are marked with different symbols and colours as explained in the figure caption. 

It should be noticed that, as we said before, the [\ion{N}{ii}] $\lambda$6583 line may be polluted with the stellar \ion{C}{ii} $\lambda$6582.9 line previous to 2005, therefore, we used the [\ion{N}{ii}] $\lambda$6548 line intensity multiplied by 3, to obtain the [\ion{N}{ii}] $\lambda$6583 intensity. The solid and dashed lines in this figure correspond to \citet{kewley:01} models and their uncertainties, respectively. These models helps to separate the regions photoionised by stars from the ones ionised by other mechanisms.
 
  In Fig. \ref{fig:bpt}  it is found that the empty symbols (corresponding to the outer nebula)  and the filled symbols (corresponding to the inner nebula) for the same epoch show a systematic different behaviour in the three BPT diagrams. Empty symbols appear always down and to the left of filled symbols, that is, they present lower values of the [\ion{O}{iii}]$\lambda$5007/H$\beta$ intensity ratio and lower values of the low-ionisation lines, relative to H$\alpha$. This is particularly noticeable for the GTC-2018 data (magenta diamonds) where the filled symbols (inner nebula) lie to the right of the Kewley's models in the cases of [\ion{N}{ii}]$\lambda$6583/H$\alpha$ and [\ion{O}{i}]$\lambda$6300/H$\alpha$, (the case of [\ion{S}{ii}]/H$\alpha$ line ratio is marginal), indicating that the ionising mechanisms in this zone is not photoionisation but a different one, like heating by shocks.  As this occurs for all the epochs, it implies that the inner nebula appears always as heated by shocks.  In the case of  [\ion{S}{ii}] lines (panel b), almost all the points are placed in the photoionisation zone. This would be related  to the radiation strength  in UV wavelength  adopted by Kewley´s models which were built with solar metallicities. In a metal-deficient environment such as in PM\,1-188 which shows low S abundance, the respective [\ion{O}{iii}]$\lambda$5007/H$\beta$ and [\ion{S}{ii}]/H$\alpha$ intensities are expected to become stronger and lower than the values indicated by Kewley´s models. Thus, the observed low [\ion{O}{iii}]$\lambda$5007/H$\beta$ and strong [\ion{S}{ii}]/H$\alpha$ values in the inner nebula could be explained by shocks rather than photoionisation.
  
  Thus, by means of BPT diagrams, we have corroborated that the inner nebula is mainly heated by shocks, while the outer nebula is photoionised.

\section{The outflow from the central star}

 We first noticed an outflow ejected from the central star in the OAN-SPM echelle REOSC spectrum obtained in 2004, with the slit position in the E-W direction (P.A. 90$\degr$). 
The outflow was clearly visible in both [\ion{N}{ii}]$\lambda$6548 and $\lambda$6583 lines and it did not appear in the H$\alpha$ line. The jet emission was more intense and extended to the blue side of the lines, being the red emission fainter. 
 
 These results were confirmed in  the high resolution spectrum obtained with the MES on 2017-06-29 (see Fig. \ref{fig:pvd}). The outflow found in the 2004 echelle REOSC spectrum appears better defined in the MES spectrum due to the better spectral resolution of MES, and to a larger exposure time.  This outflow is evidently ejected from the central star. The  spectrum obtained for H$\alpha$, in this orientation, shows an ellipse with a cavity in the centre where the expansion velocity is the highest.

 The spectrum obtained with MES at P.A. of 0$\degr$ does not show the outflow (see Fig. \ref{fig:pvd}). Interestingly the two lobes of the H$\alpha$ emission, corresponding to the outer nebula, appear tilted with a displacement  of about  15 km s$^{-1}$, being the northern zone to the blue.  This is showing that the outer nebula is not spherical as it appears in the images (and so it was classified by \citealt{sahai:11} as R,*,ib: round with an inner bubble, a visible star). 
  Our H$\alpha$ images are equal to the PV diagrams presented by \citet{guerrero:18} who claimed that  this behaviour corresponds to a barrel-like structure whose symmetry axis is tilted to the line of sight by 25$\degr$.
 
  In the N-S oriented image (P.A. 0$\degr$), the [\ion{N}{ii}] emission is strong and wide in the centre and faint and elongated towards the N-S direction, showing some faint filamentary  structure.
  
   The observed radial velocity of the object, derived from these high resolution spectra, is $+56.2\pm$1 km s$^{-1}$ and the heliocentric velocity is v$_{sys}$ = 53.9$\pm$1 km s$^{-1}$.
   
  The velocity fields obtained from the MES high resolution spectra are analysed in the next section, by means of Position-Velocity diagrams.

\section{Position Velocity diagrams}

Position-velocity diagrams (PVD) were constructed for the MES 2017 data at slit positions  P.A. 90$\degr$ and P.A. 0$\degr$, for the [\ion{N}{ii}]$\lambda$6583 and H$\alpha$ lines in order to explore the nebular kinematics. It is safe to use these lines because at this epoch the star was too faint, it was not detected and does not contaminate the nebular lines. These PVD were produced in Python, using the Astropy library. Our code reads our MES data and plots them using the proper scale plate and the radial velocity resolution. 
As the slit was placed on the central star for the two positions, P.A. 90$\degr$ and P.A. 0$\degr$,  the spatial origin is the position of the central star  and the spatial distribution takes into account a pixel size of 0.36$''$. To reconstruct the radial velocities, we used the rest wavelengths of H$\alpha$ and  [\ion{N}{ii}]$\lambda$6583  lines, relatives to the systemic velocity (v$_{\rm sys}$=53.9 km s$^{-1}$), thus we can analyse the kinematics of these lines.

The PV diagrams are presented in Fig. \ref{fig:pvd}. The line contours in each diagram represent the intensities of the lines, each line has its ad-hoc contour separation, as explained later. 

\subsection{The PV diagrams at P.A. 90\degr}
  In Fig. \ref{fig:pvd} (up, left),  we can observed the H$\alpha$ emission presenting two peaks and a cavity in the centre. The contours show a maximum expansion velocity HWHM of about 40 km s$^{-1}$ in the centre.
  
  The contours  for the H$\alpha$ PVD represent the 10\%, 20\%, 30\%, 40\%, and 50\% of the total intensity from the outer to the inner contour respectively. The 10\% contour shows a faint elongation to the blue.
  
The [\ion{N}{ii}]$\lambda$6583 PV diagram (Fig. \ref{fig:pvd},  up, right) shows a complex structure. Here the contours represent the 7\%, 10\%, 15\%, 20\%, 30\% and 40\% of the total intensity from the outer to the inner contour respectively. The structure does not exhibit the two peaks  noticed in H$\alpha$, but only one centrally concentrated bright emission (slightly blue shifted).  This zone has an expansion velocity of about 40 km s$^{-1}$. The most spectacular fact in this diagram is the evident outflow emerging from the central star,
  with velocities ranging from $-$150 to 100 km s$^{-1}$. The outflow to the blue is more extended and brighter than the red one, possibly due to a larger extinction in the back.  The outflow is visible mainly in the [\ion{N}{ii}] lines. It is only barely observed in H$\alpha$ because its position coincides with the cavity shown in H$\alpha$ in the central zone. 

\subsection{The PV diagrams at P.A. 0\degr}
\begin{figure*}
	\includegraphics[width=\columnwidth]{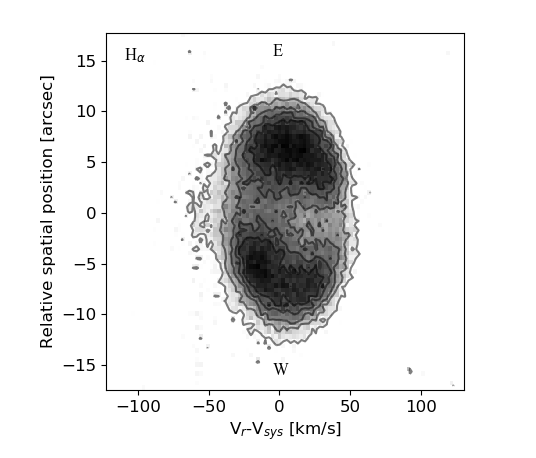}
	\includegraphics[width=\columnwidth]{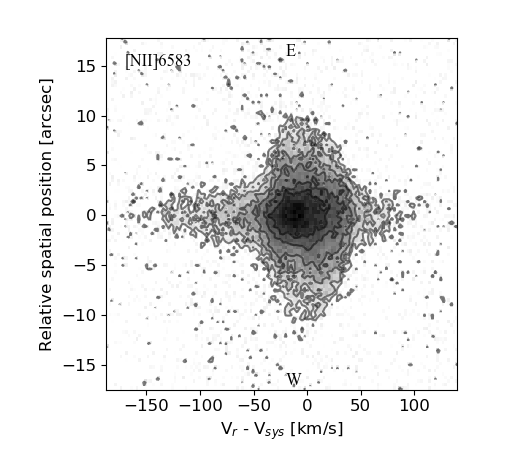}
	\includegraphics[width=\columnwidth]{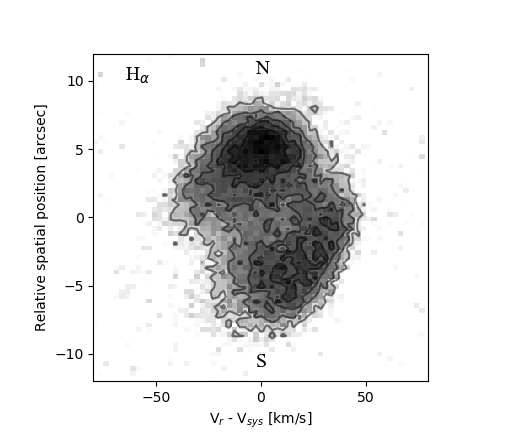}
	\includegraphics[width=\columnwidth]{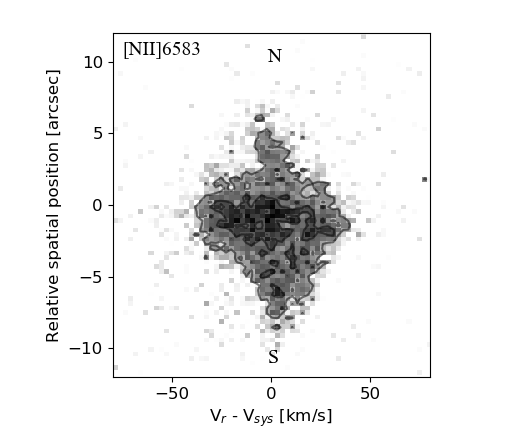}
    \caption{Position-velocity diagrams for the H$\alpha$ and [\ion{N}{ii}]$\lambda$6583 lines are presented, from the observations with MES. Upper panels present the PV diagrams with the slit oriented E-W (P.A. 90$\degr$) and lower panels, with the slit oriented N-S (P.A. 0$\degr$).}
    \label{fig:pvd}
\end{figure*}

The H$\alpha$ and [\ion{N}{ii}]$\lambda$6583 PV diagrams, for this direction, are shown in Fig. \ref{fig:pvd} down (left and right respectively). 

The contours in the H$\alpha$ PVD represent the same intensity levels as in the E-W PV diagram. In this case the H$\alpha$ lobes, corresponding to the north and south sides of the outer nebula, are displaced in velocity by about 15 km s$^{-1}$ one from the other. The northern lobe is shifted to the blue.   This behaviour indicates that the nebula is not spherical, but an open-ended (barrel-like) elongated structure. 

The [\ion{N}{ii}]$\lambda$6583 PVD in this orientation shows a concentrated  central zone with an expansion velocity 
$\leq$ 40 km s$^{-1}$. A faint filamentary structure extending from North to South, of about 20 arcsec long, corresponding with the outer zone, is noticed. No outflow is found in this orientation.

 From the PV diagrams the kinematic ages for the inner and outer nebulae can be derived.  A distance of 5.36 kpc to  PM\,1-188 as published by \citet{frew:16} was adopted. Therefore, from our H$\alpha$ image at P.A.=90$\degr$ which shows a projected nebular size of 10 arcsec  and an expansion velocity  of 40 km s$^{-1}$, we compute a kinematic nebular age  of  6,400 yr for the outer nebula. This is a lower limit if the structure is elongated and inclined with respect to the line of sight as proposed by \citet{guerrero:18}. For the inner nebula we used the [\ion{N}{ii}]$\lambda$6583 P.A.=90$\degr$ image and performed the same calculations taking into account a size of 3 arcsec and an expansion velocity of 40 km s$^{-1}$, from these data we computed a lower-limit kinematic age of 1,900 yr. A kinematic age can be determined also for the outflow by assuming that we are seeing only one pulse. This structure shows a size of 2 arcsec and an expansion velocity of 150 km s$^{-1}$, then a kinematic age equal to 335 yr is derived.

\section{Discussion and conclusions}
 PM\,1-188 is a planetary nebula with a central star of the rare spectral type [WC\,10], whose brightness has been diminishing with time, losing about 10 magnitudes in the last 40 years.  The central star seems to have had an event of the  born-again type as claimed by \citet{guerrero:18}. A few born-again central stars have been reported in the literature. The most studied cases are A\,30, A\,78, A\,58 and recently WR\,72 (\citealt{wesson:08}; \citealt{gvaramatze:20}, and references therein). They have been explained through a very late thermal pulse (VLTP) of a post-AGB star.  The main characteristics of these born-again central stars (mostly classified as [WC early] or [WC early]-PG1159 stars) and their associated PNe are the existence of an outer evolved H-rich nebula, and an inner H-deficient shell which in many cases shows knots with cometary tails extended radially. Interestingly, a binary central star was found in A\,30 \citep{jacoby:20}.
 Some of these characteristics have been found in PM\,1-188, in particular the existence of an outer low density nebula, and a compact denser inner shell showing an outflow similar to the cometary tails. However the effective temperature of its central star is much lower than that found in the mentioned born-again  stars, and such objects have not shown the decrease in brightness displayed by PM\,1-188 central star.
 
 In this work the spectrophotometric  behaviour of the ionised gas in PM\,1-188 is analysed by means of medium and high resolution spectra obtained from observations along 20 years. This allows us to study the time evolution of emission lines in the inner and outer nebulae.  Also, the time evolution of physical parameters is analysed.  
 
The outer extended nebula emits H and He lines, and faint low-ionised heavy-element (N$^+$, O$^+$, and S$^+$) lines. We found that the heavy-elements emission lines have been decreasing with time and also the electron temperature has diminished from about 13,000 K previous 2005 to less than 7,000 K in 2018.  Therefore this section of the nebula, ejected during the first time the star was in the AGB zone, has been recently cooling and recombining with time as a consequence of the fading of the central star because less and less ionising photons arrive to the outer nebula with time. A kinematic age of 6,400 yr was calculated from the projected nebular dimension, the distance and the expansion velocity. This age could be larger if the structure is elongated and inclined, as propose \citet{guerrero:18}.  

The inner nebula shows very intense lines of low ionised species such as N$^+$, O$^+$, and S$^+$. Also weak lines of O$^{++}$, Ne$^{++}$ and He$^{++}$ are found in this zone.  
It is particularly interesting the faint emission of the \ion{He}{ii}$\lambda$4686 found in this nebula.
Given that the effective temperature of the central star (35,000 -- 38,000 K) is too low and it does not emit the hard UV photons needed to ionise He twice, the He$^{++}$ should be produced in the shocked region, as predicted by the shock model presented by \citet{guerrero:18}. BPT diagrams constructed by us also indicate that the ionising mechanisms in the inner nebula is not photoionisation but a different one as heating by shocks.
 Besides the unexpected emergence of \ion{He}{ii}$\lambda$4686 in this zone, we found that the intensities of [\ion{O}{iii}]$\lambda$5007, [\ion{O}{ii}]$\lambda$3727, and [\ion{N}{ii}]$\lambda$6548,83  have increased with time in the inner nebula, corresponding with an increasing electron temperature, which has varied from 11,000 K previous 2005 to more than 14,000 K in 2018. The origin of the temperature rise in this zone is evidently the  shock mentioned above.  The kinematic age derived for this nebula, from out data is about 1,900 yr. 

 Total abundances were derived from the ionic abundances, by using ICFs by \citet{kingsburgh:94} and \citet{delgado:14}, for the inner and outer nebulae. The derived values for the inner nebula are  12+log He/H = 11.13$\pm$0.05, 12+log O/H = 7.99$\pm$0.05, log N/O = $-$0.06$\pm$0.07, log Ne/O = 0.26$\pm$0.08, log S/O = $-$0.99$\pm$0.11, and log Ar/O = $-$2.32$\pm$0.09 and they are similar for the outer nebula. 
 The nebular chemistry of PM\,1-188 is peculiar.  In comparison with the chemistry of disc non-Type I PNe, with average values  12+log He/H = 11.05$\pm$0.06, 12+log O/H = 8.69$\pm$0.20, 12+log N/H = 8.14$\pm$0.20, 12+log Ne/H = 8.10$\pm$0.15, 12+log S/H= 6.90$\pm$0.30, 12+log Ar/H = 6.38$\pm$0.30 as presented by \citet{kingsburgh:94}, it is found that PM\,1-188 is very O, S and Ar poor, and He, N and Ne rich.  Thus, the abundances of PM\,1-188 do not correspond to disc PNe even when its position locates it in the galactic disc. The low abundances of the $\alpha$ elements O, S and Ar indicate that PM1-188 was formed in a low abundance medium, while the enrichment of He, N and Ne seems a consequence of the stellar nucleosynthesis. The chemistry of PM\,1-188 is reminiscent of the chemical abundances of some halo PNe (see the analysis for DdDm-1 and other halo PNe presented by \citealt{henry:08}), in particular it is reminiscent of BoBn\,1 which is very O, S and Ar poor and shows large C, N and Ne enrichment.  \citet{otsuka:10} derived 12+log O/H = 7.74, log N/O = 0.29, log Ne/O = 0.22,  log S/O = -2.42, and  log Ar/O = -3.41, and to explain the elemental abundances in BoBn\,1 \citet{otsuka:10} proposed for this nebula a progenitor of 1 -- 1.5 M$_\odot$ initial mass star or a  0.75+1.5 M$_\odot$ binary system. To suggest a binary system as origin of PM\,1-188 is tempting considering the rare behaviour of the central star which ejected a nebula more than 6,000 yr ago and actually, after an apparent born-again episode, presents atmospheric instabilities, a second shocked younger nebula and outflows. However, so far  there is not evidence  for a binary central star. 

An outflow emerging from the central zone was found in our high resolution spectra obtained in P.A. 90$\degr$. It is bright in [\ion{N}{ii}] lines and does not appear in H$\alpha$ due to H lines show a cavity in this zone. From our position-velocity diagrams, it is found that this outflow shows a velocity from $-$150  to 100 km s$^{-1}$. The blue side is brighter and the red side  is fainter possibly due to large reddening in the backside. We also estimated the age of the outflow to be 335 yr. \citet{rechy:20} reported to have found a shell-like structure expanding at 300 km s$^{-1}$ with an age of 200 yr in PM\,1-188. This coincides very well with the outflow in direction N-S we are finding. Due to we have not obtained MES observations in other directions we can not assert that the ejection has a shell-like structure, but we do not detect such an outflow in the observations with P.A. 0$\degr$. Therefore the structure detected by \citet{rechy:20} should be a broken shell. If this structure were a complete shell, the energy required to keep it in motion should be larger than 10$^{45}$ erg, which is too large for the central star to provide it. A hydrodynamical model taking into account the detected outflow and the photoionisation by the central  object is in progress to study in detail the kinematics in the inner shell and its evolution (Rodr{\'\i}guez-Gonz\'alez et al. in prep.)

Other result obtained in this work, is the systemic radial velocity of 56$\pm$1 km s$^{-1}$ and a heliocentric velocity of 53.9$\pm$1 km s$^{-1}$ obtained from our high resolution spectra.

Since its discovery, the planetary nebula PM\,1-188 has given us many surprises and we have slowly advanced in the understanding of this object, but still several facts remain unexplained. Deeper observations in the optical and other wavelength ranges are necessary to get a deep insight of this peculiar object.

\section*{Acknowledgements}
This work is based upon observations carried out at the Observatorio Astron\'omico Nacional on the Sierra San Pedro M\'artir (OAN-SPM), Baja California, M\'exico, and at Las Campanas Observatory with the Magellan  telescope Clay. Work based on data from the GTC Archive at CAB (INTA-CSIC).
We thank the daytime and night support staff at the OAN-SPM for facilitating and helping to obtain our observations, along 20 years. J.S. Rechy-Garc{\'\i}a and V. G\'omez-Llanos are thanked for helpful discussion and support during observations. L.H.-M. is grateful to Dra. Maria del Pilar Carre\'on for the hospitality at ICN, UNAM. F.R.-E. acknowledges scholarship from CONACyT, M\'exico.
This work received partial support from DGAPA-PAPIIT IN103117 and  IN105020, UNAM. 
 \medskip
 
 Data Availability Statement: The data underlying this article will be shared on reasonable request to the corresponding author.



 \bigskip
 \bigskip


 \newpage
\begin{landscape}
\begin{table*}
\centering
\scriptsize
\caption{Fluxes of the inner nebula  in different epochs, F($\lambda$)$^a$. Data previous to 2005 include stellar emission lines.  
The column I($\lambda$)$^b$ corresponds to  the dereddened fluxes.}
\begin{tabular}{llrrrrrrrrrrrrrr}
\hline
  &   Telesc. & \multicolumn{2}{c}{OAN-2m}  & \multicolumn{2}{c}{OAN-2m} & \multicolumn{2}{c}{OAN-2m} & \multicolumn{2}{c}{OAN-2m}& \multicolumn{2}{c}{LCO-Clay}  & \multicolumn{2}{c}{OAN-2m} & \multicolumn{2}{c}{GTC}  \\
 & Spectr.&  \multicolumn{2}{c}{BCh} & \multicolumn{2}{c}{BCh} & \multicolumn{2}{c}{echelle} &  \multicolumn{2}{c}{BCh} & \multicolumn{2}{c}{LSST-2} &  \multicolumn{2}{c}{BCh} & \multicolumn{2}{c}{OSIRIS-LS}\\ 
 & date  & \multicolumn{2}{c}{2000/09} & \multicolumn{2}{c}{2002/06} & \multicolumn{2}{c}{2004/04} & \multicolumn{2}{c}{2005/05} & \multicolumn{2}{c}{2005/08} & \multicolumn{2}{c}{2017/006} &  \multicolumn{2}{c}{2018/06}\\ 
Ion & $\lambda_0$ & F($\lambda$) & I($\lambda$) & F($\lambda$) & I($\lambda$) & F($\lambda$) & I($\lambda$) &F($\lambda$) &I($\lambda$) & F($\lambda$) & I($\lambda$) & F($\lambda$) &I($\lambda$) &F($\lambda$) & I($\lambda$)
 \\ 
 \hline
[\ion{O}{ii}] & 3727 & OoR & OoR & 170.0 & 386.6(17) & OoR & OoR & OoR & OoR & OoR & OoR & 214.0 & 462.4(7) & 375.0 & 799.9(4) \\
~[\ion{Ne}{iii}] & 3869 & OoR & OoR & noisy & noisy & OoR & OoR & OoR & OoR & OoR & OoR & noisy & noisy & 30.0 & 59.5(8) \\ 
H8+\ion{He}{i} & 3889 & OoR & OoR & noisy & noisy & OoR & OoR & OoR & OoR & OoR & OoR & noisy & noisy & 8.1 & 23.5(10)  \\
H7 & 3970 & noisy & noisy & noisy & noisy & OoR & OoR & OoR & OoR & OoR & OoR & noisy & noisy & 11.7 & 18.7(10)\\ 
\ion{He}{i} & 4026 & noisy & noisy & noisy & noisy & OoR & OoR & OoR & OoR & noisy & noisy & noisy & noisy &  3.1:& 4.8:  \\ 
~[\ion{S}{ii}] & 4069 & noisy & noisy & noisy & noisy & OoR & OoR & OoR & OoR & noisy & noisy & noisy & noisy & 8.0 & 14.0(10) \\
H$\delta$ & 4102 & 19.0 & 33.6(18) & noisy & noisy & OoR & OoR & 18.0 & 31.8(10) & noisy & noisy & 23.0 & 39.8(8) & 19.0 & 32.6(6) \\
H$\gamma$ & 4340 & blend & blend & 34.0 & 51.0(15) & OoR & OoR & 35.0 & 51.6(5) & noisy & noisy & 42.0 & 61.0(5) & 32.0 & 46.2(4) \\
\ion{He}{i} & 4471 & 27.0 & 36.0(28) & 11.0 & 14.8(35) & OoR & OoR & 7.0 & 9.3(30) & noisy & noisy & 7.0 & 9.2(20) & 6.0 & 7.9(20) \\ 
\ion{He}{ii}$^c$ & 4686 & WR-b & WR-b & WR-b & WR-b & < 13 & < 14.6 & 9.0 & 10.2(30) & 10.0 & 11.3(20) & 10.0 & 11.3(20) & 12.0 & 13.5(15) \\ 
H$\beta$ & 4861 & 100.0 & 100.0(12) & 100.0 & 100.0(12) & 100.0 & 100.0(12) & 100.0 & 100.0(2) & 100.0 & 100.0(2) & 100.0 & 100.0(2) & 100.0 & 100.0(2) \\ 
~[\ion{O}{iii}] & 4959 & 26.0 & 24.4(30) & 15.0 & 14.1(35) & 9.0 & 8.5(35) & 15.0 & 14.1(25) & 83.0 & 78.1(20) & 20.0 & 18.8(25) & 28.0 & 26.4(20)\\
~[\ion{O}{iii}] & 5007 & 61.0 & 55.5(20) & 41.0 & 37.2(22) & 41.0 & 37.6(22) & 43.0 & 39.1(12) & 240.0 & 219.3(12) & 57.0 & 52.1(10) & 95.0 & 86.9(10) \\ 
~[\ion{N}{i}] & 5198 & 15.0 & 12.2(35) & 9.0 & 7.3(35) & noisy & noisy & 12.0 & 9.8(25) & noisy & noisy & 21.0 & 17.3(25) & 45.0 & 37.1(20) \\ 
~[\ion{N}{ii}] & 5755 & 9.0 & 5.7(35) & 9.0 & 5.6(35) & noisy & noisy & 6.0 & 3.8(25) & 10.4 & 6.7(25) & 19.0 & 12.2(25) & 40.0 & 25.9(10) \\
\ion{He}{i} & 5876 & 214.0 & 129.1(30) & 54.0 & 32.2(35) & 27.0 & 17.0(35) & 35.0 & 21.1(15) & 26.0 & 16.0(15) & 29.0 & 17.8(15) & 28.0 & 17.4(12) \\
~[\ion{O}{i}] & 6300 & 36.0 & 18.7(35) & 13.0 & 6.7(35) & noisy & noisy & 12.0 & 6.3(25) & 37.0 & 19.8(25) & 47.0 & 25.1(20) & 94.0 & 50.7(10) \\
~[\ion{O}{i}] & 6364 & blend & blend & blend & blend & noisy & noisy & 5.0: & 2.6: & noisy & noisy & 20.0 & 10.5(25) & 33.0 & 17.4(20) \\ 
~[\ion{N}{ii}] & 6548 & 253.0 & 121.3(18) & 196.0 & 92.2(20) & 130.0 & 66.0(20) & 143.0 & 68.6(10) & 257.0 & 127.0(8) & 299.0 &147.4(8) &545.0 & 272.0(6) \\
H$\alpha$ & 6563 & 600.0 & 286.2(14) & 603.0 & 282.2(14) & 566.0 & 286.2(14) & 600.0 & 286.2(4) & 582.0 & 286.3(4) & 569.0 & 279.2(4) & 562.0 & 279.2(4) \\ 
~[\ion{N}{ii}] & 6583 & 1279.0 & 605.9(14) & 648.0 & 301.1(14) & 392.0 & 197.0(14) & 476.0 & 225.5(4) & 1083.0 & 529.3(4) & 1009.0 & 491.7(4) & 1697.0 & 837.4(3) \\
\ion{He}{i} & 6678 & OoR & OoR & 41.0 & 18.5(25) & noisy & noisy & 18.0 & 8.3(25) & OoR & OoR & 12.0 & 5.7(25) & 10.0 & 4.8(25) \\ 
~[\ion{S}{ii}] & 6716 & OoR & OoR & 73.0 & 32.4(35) & 67.0 & 32.3(35) & 66.0 & 29.9(25) & OoR & OoR & 102.0 & 47.6(25) & 152.0 & 72.0(7) \\ 
~[\ion{S}{ii}] & 6731 & OoR & OoR & 60.0 & 26.5(35) & 58.0 & 27.9(35) & 55.0 & 24.8(25) & OoR & OoR & 99.0 & 46.1(25) & 137.0 & 64.6(8) \\
\ion{He}{i} & 7065 & OoR & OoR & 59.0 & 23.3(40) & --- & --- & 23.0 & 9.3(25) & OoR & OoR & 15.0 & 6.3(25) & 24.0 & 10.2(20) \\ 
~[\ion{Ar}{iii}] & 7136 & OoR & OoR & noisy & noisy & --- & --- & 15.0 & 5.9(25) & OoR & OoR & 15.0 & 6.1(25) & 14.0 & 5.8(20) \\ 
\ion{C}{ii} & 7236 & OoR & OoR & 372.0 & 138.5(25) & --- & --- & 113.0 & 43.1(15) & OoR & OoR & 20.0 & 7.9(25) & 13.0 & 5.2(20) \\
~[\ion{O}{ii}]+ & 7325 & OoR & OoR & 31.0 & 11.2(30) & --- & --- & OoR & OoR & OoR & OoR & 73.0 & 28.1(20) & 122.0 & 47.7(15) \\ 
\hline
  \multicolumn{2}{l}{F(H$\beta$)$^d$}  &14.9 & &20.2 & &9.46 & &8.20 & &0.458 & &5.72 && 0.727 \\
  c(H$\beta$) & & \multicolumn{2}{l}{1.08$\pm$0.05} & \multicolumn{2}{l}{1.11$\pm$0.07}  & \multicolumn{2}{l}{0.99$\pm$0.07} &  \multicolumn{2}{l}{1.08$\pm$0.04} & \multicolumn{2}{l}{1.04$\pm$0.03} &\multicolumn{2}{l}{1.04$\pm$0.04} &\multicolumn{2}{l}{1.02$\pm$0.03} \\
\hline
\multicolumn{16}{l}{OoR is Out of Range; --- indicates line not detected or not measurable.}\\
\multicolumn{16}{l}{$^a$ The errors in flux are about 2\% when F($\lambda$)/F(H$\beta$)$\geq$1,  about 10 \% when F($\lambda$)/F(H$\beta$) $\sim$ 0.5, about 20\% when F($\lambda$)/F(H$\beta$)$\leq$0.2 }\\
\multicolumn{16}{l}{$^a$ lines marked with : are very uncertain.}\\
\multicolumn{16}{l}{$^b$ The uncertainties in dereddened intensities I($\lambda$), calculated by considering the uncertainties in c(H$\beta$), are given in \% and enclosed in parenthesis.}\\
\multicolumn{16}{l}{$^c$ WR-b is Wolf-Rayet bump.}\\
\multicolumn{16}{l}{$^d$ The absolute flux in H$\beta$ is in unites of E-15 erg cm$^{-2}$ s$^{-1}$ and it is only indicative as it depends on the extraction window, the observing conditions (photometric or not), etc.}\\
\end{tabular}
\label{tab:inner-nebula}
\end{table*} 
\end{landscape}

\newpage
\begin{table*}
\caption{Fluxes of the outer nebula  in different epochs, F($\lambda$)$^a$. Data previous to 2005 can include stellar emission.  
The column I($\lambda$)$^b$ corresponds to  the dereddened fluxes.}
\begin{tabular}{llrrrrrrrrrrrr}
\hline
  &   Telesc. & \multicolumn{2}{c}{OAN-2m}  & \multicolumn{2}{c}{OAN-2m} &  \multicolumn{2}{c}{OAN-2m}& \multicolumn{2}{c}{LCO-Clay}  & \multicolumn{2}{c}{OAN-2m} & \multicolumn{2}{c}{GTC}  \\
 & Spectr.&  \multicolumn{2}{c}{BCh} & \multicolumn{2}{c}{BCh} &   \multicolumn{2}{c}{BCh} & \multicolumn{2}{c}{LSST-2} &  \multicolumn{2}{c}{BCh} & \multicolumn{2}{c}{OSIRIS-LS}\\ 
 & Date  & \multicolumn{2}{c}{2000/09} & \multicolumn{2}{c}{2002/06} &  \multicolumn{2}{c}{2005/05} & \multicolumn{2}{c}{2005/08} & \multicolumn{2}{c}{2017/06} &  \multicolumn{2}{c}{2018/06}\\ 
Ion & $\lambda_0$ & F($\lambda$) & I($\lambda$) & F($\lambda$) & I($\lambda$) &F($\lambda$) & I($\lambda$) &F($\lambda$) &I($\lambda$) & F($\lambda$) & I($\lambda$) & F($\lambda$) &I($\lambda$) \\ 
 \hline
[\ion{O}{ii}] & 3727 & OoR & OoR & 98.0 & 303.2(30) & noisy & noisy & OoR & OoR & noisy & noisy & 21.0 & 41.0(20)\\
H9 & 3835 & OoR & OoR & noisy & noisy & noisy & noisy & noisy & noisy & noisy & noisy & 4.3: & 8.0:\\
H8+\ion{He}{i} & 3889 & OoR & OoR & noisy & noisy & noisy & noisy & noisy & noisy & noisy & noisy & 12.2 & 27.2(15)\\
H7 & 3970 & OoR & OoR & noisy & noisy & noisy & noisy & noisy & noisy & 14.3 & 24.6(20) & 11.0 & 13.9(20)\\
\ion{He}{i} & 4026 & OoR & OoR & noisy & noisy & noisy & noisy & noisy & noisy & noisy & noisy & 2.7: & 5.2:\\
~[\ion{S}{ii}] & 4068 & OoR & OoR & noisy & noisy & noisy & noisy & noisy & noisy & noisy & noisy & 3.3: & 4.4:\\
H$\delta$ & 4102 & OoR & OoR & noisy & noisy & 15.9 & 30.0(10) & 27.0 & 46.7(10) & 23.9 & 38.1(10) & 17.0 & 27.4(8)\\
H$\gamma$ & 4340 & 49.7 & 75.1(20) & 25.7 & 44.5(20) & 33.0 & 50.9(10) & 35.1 & 50.9(10) & 41.5 & 57.0(10) & 36.0 & 49.8(8)\\
\ion{He}{i} & 4471 & blend & --- & 11.1 & 16.6(40) & 6.0 & 8.3(25) & 3.4: & 4.5: & 6.3 & 8.0(25) & 7.3 & 9.3(20)\\
H$\beta$ & 4861 & 100.0 & 100.0(12) & 100.0 & 100.0(12) & 100.0 & 100.0(2) & 100.0 & 100.0(2) & 100.0 & 100.0(2) & 100.0 & 100.0(2)\\
\ion{He}{i} & 4922 & 14.2 & 13.6(35) & 3.3 & 3.1(45) & 4.5 & 4.3(25) & --- & --- & --- & --- & 2.2 & 2.1(25)\\
~[\ion{O}{iii}] & 4959 & 20.6 & 19.3(30) & 14.2 & 13.0(30) & 7.2 & 6.7(20) & noisy & noisy & noisy & noisy & noisy & noisy\\
~[\ion{O}{iii}] & 5007 & 29.3 & 26.5(25) & 41.3 & 36.2(25) & 17.1 & 15.4(15) & 15.2 & 13.9(15) & 6.6 & 6.1(20) & 3.7 & 3.4(25)\\
\ion{He}{i} & 5016 & 17.1 & 15.4(40) & 6.1 & 5.3(45) & 6.7 & 6.0(25) & noisy & noisy & noisy & noisy & 2.8 & 2.6(25)\\
~[\ion{N}{i}] & 5198 & 9.2 & 7.4(35) & 8.5 & 6.4(35) & 5.9 & 4.7(25) & noisy & noisy & 4.5 & 3.8(25) & 3.9 & 3.3(25)\\
~[\ion{N}{ii}] & 5755 & 8.1 & 5.0(35) & 8.0 & 4.2(35) & 5.8 & 3.5(25) & 4.0 & 2.6(25) & 1.2 & 0.8(30) & 0.8 & 0.6(25)\\
\ion{He}{i} & 5876 & 108.1 & 63.2(30) & 53.7 & 26.3(30) & 36.6 & 20.8(10) & 17.3 & 10.7(10) & 29.8 & 19.7(10) & 30.7 & 20.1(10)\\
~[\ion{O}{i}] & 6300 & 28.4 & 14.2(35) & 14.1 & 5.6(40) & 9.6 & 4.6(30) & noisy & noisy & 6.1 & 3.6(30) & 3.8 & 2.2(25)\\
~[\ion{O}{i}] & 6364 & 28.5 & 13.9(35) & 7.8: & 3.0: & 4.7: & 2.2: & noisy & noisy & noisy & noisy & 1.4: & 0.8: \\
~[\ion{N}{ii}] & 6548 & 132.0 & 60.5(20) & 174.0 & 61.7(20) & 125.7 & 55.4(10) & 79.9 & 39.5(10) & 52.4 & 28.7(10) & 69.6 & 37.7(8)\\
H$\alpha$ & 6563 & 618.9 & 282.2(13) & 802.0 & 282.4(13) & 653.3 & 286.2(3) & 582.0 & 286.32(3) & 537.0 & 293.2(3) & 544.3 & 293.1(3)\\
~[\ion{N}{ii}] & 6583 & 674.2 & 305.1(14) & 708.2 & 246.9(14) & 417.1 & 181.4(4) & 272.0 & 132.9(4) & 194.0 & 105.3(4) & 214.5 & 114.9(3)\\
\ion{He}{i} & 6678 & 67.6 & 29.6(35) & 41.0 & 13.6(35) & 17.7 & 7.4(20) & OoR & OoR & 9.8 & 5.2(20) & 9.5 & 5.0(15)\\
~[\ion{S}{ii}] & 6716 & 73.9 & 31.9(40) & 73.0 & 24.3(35) & 60.4 & 25.0(25) & OoR & OoR & 37.4 & 19.6(25) & 38.9 & 20.1(10)\\
~[\ion{S}{ii}] & 6731 & 71.4 & 30.7(35) & 60.0 & 19.9(40) & 49.6 & 20.4(30) & OoR & OoR & 29.9 & 15.6(30) & 29.1 & 15.0(15)\\
\ion{He}{i} & 7065 & OoR & OoR & 59.0 & 16.5(35) & 15.4 & 5.6(20) & OoR & OoR & 5.2: & 2.5: & 4.2: & 2.0:\\
~[\ion{Ar}{iii}] & 7136 & OoR & OoR & noisy & noisy & 5.6: & 2.0: & OoR & OoR & 5.8: & 2.7: & 1.0: & 0.5:\\
\ion{C}{ii} & 7236 & OoR & OoR & 371.7 & 95.6(20) & 68.2 & 23.3(15) & OoR & OoR & 4.9: & 2.2: & 1.4: & 0.6:\\
~[\ion{O}{ii}]+ & 7325 & OoR & OoR & 31.2 & 7.7(40) & OoR & OoR & OoR & OoR & 3.3 & 1.5(30) & 1.8 & 0.8(25)\\
 \hline
 \multicolumn{2}{l}{F(H$\beta$)$^c$ }   & 2.83 &  & 2.02 &  & 2.46 &  & 0.396 &  & 0.75 &  & 0.13 & \\
c(H$\beta$) &  & \multicolumn{2}{l}{1.15$\pm$0.08} &  \multicolumn{2}{l}{1.52$\pm$0.10} &  \multicolumn{2}{l}{1.20$\pm$0.10} &  \multicolumn{2}{l}{1.04$\pm$0.05} &  \multicolumn{2}{l}{0.88$\pm$0.08} &  \multicolumn{2}{l}{0.90$\pm$0.04}  \\
\hline
\multicolumn{9}{l}{OoR is Out of Range; --- indicates line not detected or not measurable.}\\
\multicolumn{14}{l}{$^a$ The errors in flux are about 2-3\% when F($\lambda$)/F(H$\beta$)$\geq$1,  about 10 \% when F($\lambda$)/F(H$\beta$) $\sim$ 0.5, about 20\% when F($\lambda$)/F(H$\beta$)$\leq$0.2, }\\
\multicolumn{14}{l}{$^a$ lines marked with : are very uncertain.}\\
\multicolumn{14}{l}{$^b$ The errors in I($\lambda$), taking into account the uncertainties in c(H$\beta$), are given in \% enclosed in parenthesis.}\\
\multicolumn{14}{l}{$^c$ The absolute flux in H$\beta$ is in units of E-14 erg cm$^{-2}$ s$^{-1}$ and it is only indicative as it depends on the extraction window, the observing conditions, etc.}\\
\end{tabular}
\label{tab:outer-nebula}
\end{table*}

\begin{landscape}
\begin{table*}

\setlength\tabcolsep{4pt}

\centering
	\caption{Physical conditions, ionic and total abundances for the central and outer nebulae, obtained with PyNeb, for all the epochs}
	\label{tab:abundance_table}
	\scriptsize
\begin{tabular}{lcccccccccccc}

\hline
Date (yy-mm) & 00-09 & 02-06 & 05-05 & 05-08 & 17-06 & 18-06 & 00-09 & 02-06 & 05-05 & 05-08 & 17-06 & 18-06\\
Zone& Inner & Inner & Inner & Inner & Inner & Inner & Outer & Outer & Outer & Outer & Outer & Outer  \\ \hline
n$\mathrm{_e}$ [\ion{S}{ii}] (cm$^{-3}$) & 100* & 188$\pm$50 &  203$\pm$50 & 100* & 479$\pm$50 & 348$\pm$50 & 469$\pm$108 & 191$\pm$78 & 186$\pm$73 & 100* & 140$\pm$61 & 67$\pm$45 \\
T$\mathrm{_e}$ [\ion{N}{ii]} ($10^4$ K) & 1.03$\pm$0.07 & 1.15$\pm$0.08 & 1.10$\pm$0.07 & 1.08$\pm$0.08 & 1.33$\pm$0.12 & 1.44$\pm$0.03 & 1.33$\pm$0.11 & 1.21$\pm$0.09 & 1.16$\pm$0.08 & 1.19$\pm$0.09 & 0.85$\pm$0.06 & 0.69$\pm$0.03 \\ 
T$\mathrm{_e}$ [\ion{S}{ii}] ($10^4$ K) & --- &  ---&  --- & --- &  --- & 1.54$\pm$0.06 & --- & --- & --- & --- & --- & --- \\
T$\mathrm{_e}$ [\ion{O}{ii}] ($10^4$ K) & --- & 1.31$\pm$0.23 & --- & --- & 2.03$\pm$0.28 & 2.23$\pm$0.25 & --- & 1.19$\pm$0.10 & --- & --- & --- & 1.08$\pm$0.05\\ 

\hline\hline
He$^+$/H$^+$ & --- & --- &  0.15$\pm$0.02 & 0.12$\pm$0.02 &  0.13$\pm$0.02 & 0.12$\pm$0.01 &  --- & --- & 0.15$\pm$0.01 & 0.08$\pm$0.01 & 0.14$\pm$0.02 & 0.14$\pm$0.02\\
He$^{++}$/H$^+$ (10$^{-2}$)& --- & --- &  0.84$\pm$0.09 &  0.93$\pm$0.20 & 0.96$\pm$0.21 & 1.16$\pm$0.18 & --- & --- & --- & --- & --- & --- \\  
ICF(He$^+$+He$^{++}$) & --- & --- & 1.00 & 1.00 & 1.00 & 1.00 & --- & --- & 1.00 & 1.00 & 1.00 & 1.00\\
\hline 
O$^+$/H$^+$ (10$^{-5}$) &--- & 8.88$\pm$5.44 & --- & --- & 6.15$\pm$1.60 & 8.24$\pm$0.30 & --- & 5.77$\pm$3.19 & --- & --- & --- & 10.05$\pm$0.90\\ 
O$^{++}$/H$^+$ ($10^{-5}$) & 2.07$\pm$0.10 & 0.89$\pm$0.06 & 1.04$\pm$0.03 & 6.19$\pm$0.53 & 0.78$\pm$0.06 & 0.97$\pm$0.09 & 0.61$\pm$0.03 & 0.72$\pm$0.04 & 0.38$\pm$0.02 & 0.28$\pm$0.02 & 0.39$\pm$0.03 & 0.53$\pm$0.02\\
ICF(O$^+$+O$^{++}$) & --- & 1.00 & --- & --- & 1.04$\pm$0.15 & 1.05$\pm$0.11 & --- & 1.00 & --- & --- & --- & 1.00 \\ 
\hline
N$^+$/H$^+$ ($10^{-5}$) & 6.92$\pm$0.22 & 3.97$\pm$0.12 & 3.28$\pm$0.23 & 6.39$\pm$0.70 & 4.52$\pm$0.54 & 7.10$\pm$0.10 & 1.87$\pm$0.10 & 2.36$\pm$0.10 & 2.31$\pm$0.16 & 1.58$\pm$0.11 & 2.71$\pm$0.30 & 6.97$\pm$0.58 \\
ICF(N$^+$) & --- & 1.10$\pm$0.91 & --- & --- & 1.17$\pm$0.44 & 1.18$\pm$0.14 & --- & 1.12$\pm$0.83 & --- & --- & --- & 1.05$\pm$0.12 \\
\hline
S$^+$/H$^+$ ($10^{-6}$) & --- & 1.09$\pm$0.22 &  1.13$\pm$0.11 & --- & 1.35$\pm$0.10 & 1.64$\pm$0.06 & 0.91$\pm$0.18 & 0.73$\pm$0.17 & 0.81$\pm$0.12 & --- & 1.37$\pm$0.13 & 2.53$\pm$0.12 \\
ICF(S$^+$+S$^{++}$) & --- & 1.00$\pm$0.07 & --- & --- & 1.00$\pm$0.05 & 1.00$\pm$0.01 & --- & 1.00$\pm$0.07 & --- & --- & --- & 1.00$\pm$0.01\\
\hline
Ne$^{++}$/H$^+$ ($10^{-5}$)&    --- & --- & --- & --- &  --- & 1.77$\pm$0.02 & --- & --- & --- & --- & --- & ---\\
ICF(Ne$^{++}$) & --- & --- & --- & --- & --- & 10.00$\pm$1.43 & --- &  --- & --- & --- & --- & --- \\
\hline
Ar$^{++}$/H$^+$ (10$^{-7}$) & --- & --- & 4.31$\pm$0.45 & --- & 3.07$\pm$0.24 & 2.47$\pm$0.40 & --- & --- & 1.28$\pm$0.47 & --- & 3.64$\pm$1.21 & 1.13$\pm$0.41\\
ICF(Ar$^{++}$) & --- & --- & 1.87 & --- & 1.87 & 1.87 & --- & --- & 1.87 & --- & 1.87  & 1.87 \\ \hline
\hline
He/H & --- & --- & 11.21$\pm$0.06 & 11.11$\pm$0.07 & 11.14$\pm$0.06 & 11.13$\pm$0.05 & --- & --- & 11.19$\pm$0.04 & 10.91$\pm$0.05 & 11.14$\pm$0.05 & 11.13$\pm$0.05\\
O/H & --- & 7.99$\pm$0.24 & --- & --- & 7.86$\pm$0.12 & 7.99$\pm$0.05 & --- & 7.81$\pm$0.21 & --- & --- & --- & 8.04$\pm$0.04 \\  
N/H & --- & 7.64$\pm$0.36 & --- & --- & 7.72$\pm$0.17 & 7.92$\pm$0.05 & --- & 7.42$\pm$0.32 & --- & --- & --- & 7.87$\pm$0.06 \\  
Ne/H & --- & --- & --- & --- & --- & 8.25$\pm$0.06 & --- &  --- & --- & --- & --- & --- \\
S/H & --- & 6.82$\pm$0.10 & --- & --- & 6.92$\pm$0.10 & 7.00$\pm$0.10 & --- & 6.65$\pm$0.10 & --- & --- & --- & 7.18$\pm$0.10\\
Ar/H & --- & --- &5.91$\pm$0.05 & --- & 5.75$\pm$0.03 & 5.66$\pm$0.07 & --- & --- & 5.38$\pm$0.16 & --- & 5.83$\pm$0.14  & 5.33$\pm$0.16 \\ \hline
N/O & --- & $-$0.35$\pm$0.43 & --- & --- & $-$0.13$\pm$0.21 & $-$0.06$\pm$0.07 &  --- & $-$0.39$\pm$0.39 & --- & --- & --- & $-$0.18$\pm$0.07 \\  
Ne/O & --- & --- & --- & --- & --- & 0.26$\pm$0.08 & --- &  --- & --- & --- & --- & --- \\
S/O & --- & $-$1.17$\pm$0.26 & --- & --- & $-$0.94$\pm$0.15 & $-$0.99$\pm$0.11 & --- & $-$1.16$\pm$0.24 & --- & --- & --- & $-$0.87$\pm$0.11\\
Ar/O & --- & --- & --- & --- & $-$2.11$\pm$0.12 & $-$2.32$\pm$0.09 & --- & --- & --- & --- & ---  & $-$2.73$\pm$0.21 \\
\hline \\
\multicolumn{13}{l}{$*$ Adopted values.}\\ 
\multicolumn{13}{l}{$^a$ Total abundances in 12+log X/H scale.}\\ 
\multicolumn{13}{l}{$^b$ Solar abundances adopted from \citep{asplund:09} are 12+log He/H=10.93, 12+log O/H=8.69, 12+log C/H=8.43, 12+log N/H= 7.83, 12+log Ne/H= 7.93, 12+log Ar/H= 6.40,  12+log S/H=7.12}\\ 
\end{tabular}
\end{table*}
\end{landscape}



\appendix
\section{Atomic parameters used in PyNeb calculations and the Ionisation Correction Factors}

ICFs used for the total abundances calculation are listed next. Those marked with \citetalias{kingsburgh:94} correspond to the expressions from \citet{kingsburgh:94} and those marked \citetalias{delgado:14} correspond to the expressions given by \citet{delgado:14}.

\begin{itemize}

\item $\mathrm{ \frac{He}{H} = \frac{He^{+}}{H^{+}} }$. If He$^{++}$ is detected $\mathrm{ \frac{He}{H} = \frac{He^{+} + He^{++}}{H^{+}} }$.

\item $\mathrm{ \frac{O}{H} = ICF(O) \times \frac{O^+ + O^{++}}{H^+}}$. $\mathrm{ICF(O) = 1}$ if no He$^{++}$ is detected. \\ Otherwise, $\mathrm{ICF(O)}$ is given by the equation (12) in \citetalias{delgado:14}. 

\item $\mathrm{ \frac{N}{H} = ICF(N) \times \frac{N^+}{H^+}}$. $\mathrm{ICF(N) = \frac{O}{O^+}}$ \citepalias{kingsburgh:94}.

\item $\mathrm{ \frac{Ar}{H} = ICF(Ar) \times \frac{Ar^{++}}{H^{+}} }$. $\mathrm{ICF(Ar) = 1.87 }$ \citepalias{kingsburgh:94}.

\item $\mathrm{ \frac{Ne}{H} =  ICF(Ne) \times \frac{Ne^{++}}{H^{+}} }$. $\mathrm{ICF(Ne) = \frac{O}{O^{++}} }$ \citepalias{kingsburgh:94}.

\item $\mathrm{ \frac{S}{H} = ICF(S) \times \frac{S^{+} + S^{++}}{H^{+}} }$. $\mathrm{ICF(S) = \left[ 1 - \left( 1 - \frac{O^+}{O} \right)^3 \right]^{-1/3}}$.

As no S$^{+2}$ is detected, it can be estimated through the expression: $\mathrm{\frac{S^{++}}{S^+} = 4.677 + \left( \frac{O^{++}}{O^+}  \right)^{0.433} } $ \citepalias{kingsburgh:94}.

\end{itemize}
             
 \bigskip
             
 \begin{table}
\centering
	\caption{\bf Atomic parameters used in PyNeb calculations}
\begin{tabular}{lll}
Ion & Transition probabilities & Collisional strenghts \\
\hline
N$^+$ & \citet{froese:04}  & \citet{tayal:11}\\
O$^+$ & \citet{froese:04} & \citet{kisielius:09}\\
O$^{++}$ & \citet{froese:04} & \citet{storey:14} \\
         & \citet{storey:00} &  \\
Ne$^{++}$ & \citet{galavis:97} & \citet{McLaughlin:00} \\
S$^+$ & \citet{podobedova:09} & \citet{tayal:10} \\
Ar$^{++}$ & \citet{mendoza:83} & \citet{galavis:95} \\
         & \citet{kaufman:86}. &           \\
 \hline
Ion & \multicolumn{2}{c}{Effective recombination coefficients} \\
\hline
H$^+$ & \multicolumn{2}{c}{\citet{storey:95}} \\
He$^+$ & \multicolumn{2}{c}{\citet{porter:12,porter:13}} \\
He$^{++}$ & \multicolumn{2}{c}{\citet{storey:95} }  \\
\hline
\end{tabular}
\label{tab:atomic-parameters}
\end{table}


\bsp	
\label{lastpage}
\end{document}